\documentclass[preprint,review,12pt]{elsarticle}
\pdfoutput=1
\usepackage{algorithm,algorithmicx,algpseudocode}
\usepackage[]{hyperref}
\usepackage{amssymb}
\usepackage{amsthm}
\usepackage{amsmath}
\usepackage{multirow}
\usepackage{booktabs}
\usepackage{graphicx}
\usepackage{epsfig}
\usepackage{epstopdf}
\usepackage{pdflscape}
\usepackage{color}
\usepackage{appendix}

\usepackage{amsthm}
\usepackage{amsmath}
\usepackage[english]{babel}
\usepackage{subfigure}
\usepackage{threeparttable}
\usepackage{amssymb}

\usepackage{graphicx}
\usepackage{subfigure}
\usepackage{float}
\usepackage{tabularx}
\usepackage{graphicx}
\usepackage{threeparttable}
\usepackage[figuresright]{rotating}
\allowdisplaybreaks

\theoremstyle{definition}

\theoremstyle{remark}

\journal{XXX}

\begin{document}
\begin{frontmatter}

\title{A six-point neuron-based ENO (NENO6) scheme for compressible fluid dynamics}
\author[a]{Yue Li}
\ead{yue06.li@tum.de}

\author[b,c,d]{Lin Fu\corref{cor1}}
\ead{linfu@ust.hk}
\cortext[cor1]{Corresponding author.}

\author[a]{Nikolaus A. Adams}
\ead{nikolaus.adams@tum.de}

\address[a]{Technical University of Munich, School of Engineering and Design, Chair of Aerodynamics and Fluid Mechanics, Boltzmannstr. 15, 85748 Garching bei München, Germany}

\address[b]{Department of Mechanical and Aerospace Engineering, The Hong Kong University of Science and Technology, Clear Water Bay, Kowloon, Hong Kong}
\address[c]{Department of Mathematics, The Hong Kong University of Science and Technology, Clear Water Bay, Kowloon, Hong Kong}
\address[d]{Shenzhen Research Institute, The Hong Kong University of Science and Technology, Shenzhen, China}

\begin{abstract}
In this work, we introduce a deep artificial neural network (ANN) that can detect locations of discontinuity and build a six-point ENO-type scheme based on a set of smooth and discontinuous training data. While a set of candidate stencils of incremental width is constructed, the ANN instead of a classical smoothness indicator is deployed for an ENO-like sub-stencil selection. A convex combination of the candidate fluxes with the re-normalized linear weights forms the six-point neuron-based ENO (NENO6) scheme. The present methodology is inspired by the work [Fu et al., Journal of Computational Physics 305 (2016): 333-359] where contributions of candidate stencils containing discontinuities are removed from the final reconstruction stencil. The binary candidate stencil classification is performed by a well-trained ANN with high fidelity. The proposed framework shows an improved generality and robustness compared with other ANN-based schemes. The generality and performance of the proposed NENO6 scheme are demonstrated by examining one- and two-dimensional benchmark cases with different governing conservation laws and comparing to those of WENO-CU6 and TENO6-opt schemes.  
\end{abstract}

\begin{keyword}
TENO, ENO, artificial neural network, high-order scheme, shockwave.
\end{keyword}

\end{frontmatter}


%
\section{Introduction}

High-order and high-resolution shock-capturing schemes are essential numerical methods to solve compressible fluid problems, which may involve discontinuities and broadband flow scales \cite{pirozzoli2011numerical}\cite{shu2009high}\cite{fu2021shock}\cite{griffin2021velocity}. Lack of solution regularity invokes the Gibbs phenomenon and may lead to nonlinear instability.
Among all the concepts proposed in the past decades to cope with this issue \cite{von1950}\cite{Jameson1994}\cite{Harten1983}\cite{Liu1994}\cite{Jiang1996}\cite{fu2016family}, the family of essentially non-oscillatory (ENO) schemes probably are the most popular methods. The key concept of the ENO-type schemes is that candidate stencils on which the solution reconstruction is insufficiently smooth are identified such that their contributions can be suppressed in the final construction. For ENO schemes \cite{harten1987uniformly}, only the smoothest candidate stencil is selected, whereas, in weighted ENO (WENO) \cite{Liu1994} schemes, all candidate stencils are deployed according to their renormalized optimal weights. WENO schemes outperform ENO schemes by restoring the optimal high-order accuracy for smooth flows asymptotically.
In contrast to WENO-like smooth convex combination, the weighting strategy in targeted ENO (TENO) schemes \cite{fu2016family}\cite{fu2017targeted}\cite{fu2018new}\cite{fu2019low}\cite{fu2022efficient}\cite{takagi2022novel}\cite{fu2021very}\cite{ji2022class} either deploys a candidate stencil with its optimal linear weight or discards it completely when crossed by a discontinuity. In this way, the performance of ENO-type schemes, e.g., TENO and WENO schemes, depends on not only the smoothness measurements of candidate stencils but also on the built-in nonlinear weighting strategies. 

Since the performance of the ENO-family schemes relies on both the smoothness indicators for candidate stencils and the subsequent nonlinear weighting strategy, many variants of WENO and TENO schemes have been developed in the past years. In terms of the smoothness indicators, the standard WENO-JS \cite{Jiang1996} scheme deploys the $L_2$ norm of the derivatives of the reconstructed candidate polynomials. Through developing an approximation to derivatives with higher accuracy, a modified smoothness indicator based on the $L_1$ norm enables the WENO-NS \cite{ha2013improved} scheme not only preserve the convergence order at critical points but also predict the small-scale fluctuations better.
By introducing a new variant of smoothness indicator, the slight post-shock oscillations induced by the classical WENO schemes are effectively suppressed such that the steady state solution can be obtained \cite{zhang2007new}. Based on the criterion that smoothness indicators should be constant for a specific kind of smooth functions, e.g., the sine functions, a new smoothness indicator on four-point stencil is constructed \cite{wu2020smoothness}, which features a more succinct form and takes less floating point operations. The spectral properties of the seven-point WENO scheme with such smoothness indicators can recover to those of the underlying linear scheme for monochromatic waves.    
Concerning the nonlinear weighting strategy, the WENO-M \cite{Henrick2005} and WENO-Z \cite{Borges2008}\cite{don2013accuracy} schemes have been developed to restore the optimal accuracy order near critical points, where the classical WENO-JS \cite{Jiang1996} scheme suffers from order degeneration. To reduce the numerical dissipation of the fifth-order WENO scheme, an adaptive central-upwind sixth-order WENO-CU6 \cite{hu2010adaptive} scheme is proposed by introducing the contribution of an additional downwind stencil. More recently, the adaptive dissipation control strategy \cite{fu2019LES}\cite{fu2018new}\cite{fu2018improved} is introduced to further enhance the spectral resolution of standard TENO \cite{fu2016family}\cite{fu2017targeted}\cite{fu2019very}\cite{fu2019low}\cite{li2021low}\cite{fu2019hybrid} schemes for small-scale turbulence structures.

However, these standard and improved smoothness indicators are designed at a certain level of empiricism. Although efforts have been made to develop advanced smoothness indicators and weighting strategies, there still remains a necessity to develop an optimal candidate stencil selection strategy for the ENO-family schemes. As data-driven methods can provide an input-output mapping with a high degree of complexity, it is potential to replace the functions of classical polynomial-based smoothness indicators and even the weighting strategy for better performance. Recently, data-driven methods have been incorporated into the development of shock-capturing schemes, e.g. to function as a discontinuity indicator \cite{beck2020neural} \cite{monfort2017deep}\cite{ray2018artificial}, and to enhance the performance of specific discretization schemes \cite{bar2019learning} \cite{stevens2020enhancement}. 
An example of shock detection based on neural network is presented in \cite{ray2018artificial}, where the network is trained to serve as a troubled-cell indicator for constructing the hybrid shock-capturing scheme in the framework of Runge–Kutta discontinuous Galerkin (RKDG) method, and an extension to two-dimensional problems is given in \cite{ray2019detecting}. Morgan et al. \cite{morgan2020machine} propose a multilayer perceptron (MLP) based shock detector for a high-order finite-element method. However, as these efforts are taken to train a data-driven indicator to control the switch between different schemes, the performance of the resultant shock-capturing schemes is more dominated by the component linear and nonlinear schemes rather than the indicator itself. As an alternative approach, Stevens and Colonius \cite{stevens2020enhancement} and Bar-Sinai et al. \cite{bar2019learning} deploy neural networks to modify the coefficients of the baseline polynomial-based high-order schemes directly. It is shown in \cite{stevens2020enhancement} that the resulting WENO-NN scheme fails to preserve the optimal accuracy order and generates strong overshoots for under-resolved simulations. Moreover, the data-driven scheme coefficients for the spatial derivatives may be equation-specific and thus lack the generality and robustness \cite{bar2019learning}. More recently, Jung and Kwon \cite{jung2021flux} show that the neural network can function as a numerical model, as long as the training database is well normalized and constructed.

In this work, we propose an alternative paradigm based on ANN to replace the smoothness indicators as well as the ENO-like stencil selection in a six-point TENO6-opt scheme \cite{fu2016family}. An artificial neural network architecture is adopted to provide a nonlinear input-output mapping with a high degree of complexity such that the ENO-like stencil selection can be mimicked. This mapping is conducted through grouped neuron layers, which process data from an input of local flow field information and output the binary selection of candidate stencils based on their inherent smoothness. After training the neural network with a properly constructed database, it is able to predict the ENO-like binary stencil selection for scaled samples, which are not present in the training dataset. Similarly to the standard TENO schemes, the final high-order reconstruction is assembled by the convex combination of the polynomial-based candidate fluxes based on the binary selection indicated by the neural network, leading to the new NENO6 scheme. Compared with other ANN-based schemes, NENO6 shows an improved generality, the shock-capturing capability, and the numerical robustness.

The rest of this paper is organized as follows. In Section 2, the concept of TENO6-opt scheme is briefly reviewed. In Section 3, a short summary on the framework of the neural network is first given. Then the training data and the network architecture used for developing the NENO6 scheme are discussed in detail. In Sections 4, the performance of the proposed NENO6 scheme is demonstrated by conducting a set of critical benchmark simulations governed by different conservation laws. In Section 5, concluding remarks are given.

\section{Concepts of six-point TENO6-opt scheme for hyperbolic conservation laws}

To facilitate the presentation, we consider the one-dimensional scalar hyperbolic conservation law 
\begin{equation}
\label{eq:conservationlaw}
\frac{\partial u }{\partial t} + \frac{\partial}{\partial x}f(u) = 0 ,
\end{equation}
where $u$ and $f$ denote the conservative variable and the flux function, respectively. The characteristic signal speed is assumed to be positive with $\frac{\partial f(u)}{\partial u} > 0$. 

For a uniform Cartesian mesh with cell centers ${x_i} = i\Delta x$ and cell interfaces ${x_{i+1/2}} = {x_i} + {\Delta x}/{2}$, the finite-difference spatial discretization results in a set of ordinary differential equations (ODE) as
\begin{equation}
\label{eq:ordinaryEq}
\frac{d{u_i(t)}}{dt} =  - \frac{\partial f}{\partial x}\left| {_{x = {x_i}}} \right., \text{ } i = 0,\cdots,n,
\end{equation}
where ${u_i(t)}$ denotes a grid-function approximation of the point value $u(x_i,t)$. The semi-discretization with  Eq.~(\ref{eq:ordinaryEq}) can be further approximated by a conservative finite-difference scheme as
\begin{equation}
\label{eq:implicitEq}
\frac{d{u_i}}{dt} =  - \frac{1}{\Delta x}({h_{i + 1/2}} - {h_{i - 1/2}}) ,
\end{equation}
where the primitive function $h(x)$ is implicitly defined by
\begin{equation}
\label{eq:definitionEq}
f(x) = \frac{1}{\Delta x}\int_{x - \Delta x/2}^{x + \Delta x/2} {h(\xi )d\xi },
\end{equation}
and ${h_{i \pm 1/2}} = h({x_i} \pm {{\Delta x}}/{2})$. A high-order approximation of $h(x)$ at the cell interface has to be reconstructed from the cell-averaged values of $f(x)$ at the cell centers to complete the discretization. Eq.~(\ref{eq:implicitEq}) can be approximately written as
\begin{equation}
\label{eq:approximateEq}
\frac{d{u_i}}{dt} \approx  - \frac{1}{\Delta x}({\widehat f_{i + 1/2}} - {\widehat f_{i - 1/2}}) ,
\end{equation}
where ${\hat f_{i \pm 1/2}}$ denotes the high-order numerical fluxes and can be computed from a convex combination of $M$ candidate stencil fluxes
\begin{equation}
\label{eq:convex}
\widehat f_{i + 1/2} = \sum\limits_{k = 0}^{M-1} {w_k} \widehat f_{k,i + 1/2} ,
\end{equation}
where $w_k$ denotes the weight for each candidate flux. To achieve global order accuracy with the full stencil $K$ ($K=6$ in this work), a $({r_k} - 1)$-degree polynomial is assumed as the distribution function on each candidate stencil as
\begin{equation}
\label{eq:approximatepolynomial}
h(x) \approx {\hat f_k}(x) = \sum\limits_{l = 0}^{{r_k} - 1} {{a_{l,k}}} {x^l},
\end{equation}
where ${r_k}$ denotes the point number of candidate stencil $k$. After substituting Eq.~(\ref{eq:approximatepolynomial}) into Eq.~(\ref{eq:definitionEq}) and evaluating the integral functions at the stencil nodes, the coefficients ${a_{l,k}}$ are determined by solving the resulting system of linear algebraic equations.

In the following, we recall the essential elements of the TENO6-opt \cite{fu2016family}\cite{fu2017targeted} scheme.

\subsection{Smoothness indicators of TENO6-opt}
\begin{figure}[htbp]
\centering
\subfigure{\includegraphics[width=0.8\textwidth]{ 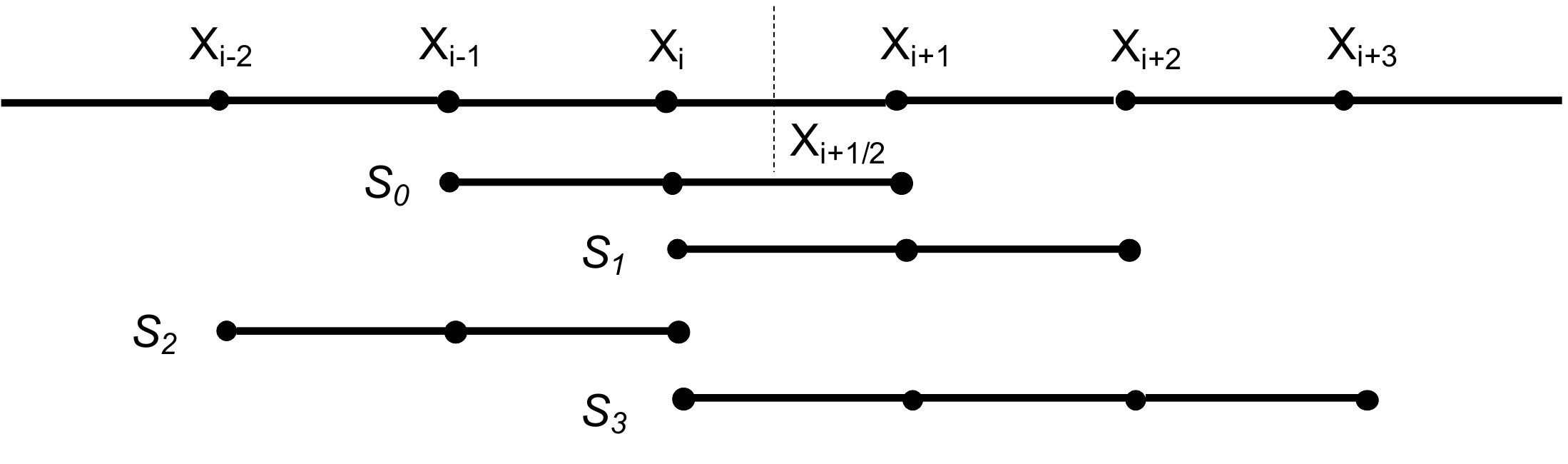}}
\caption{Candidate stencils with incremental width towards the six-point TENO6-opt reconstruction.}
\label{Fig:incremental_stencil}
\end{figure}

Different from WENO schemes, TENO6-opt scheme is constructed from a set of candidate stencils with incremental width \cite{fu2016family}, as shown in Fig.~\ref{Fig:incremental_stencil}.

To effectively isolate discontinuities from smooth regions, smoothness indicators with strong scale-separation capability are given as \cite{fu2016family}
\begin{equation}
\label{eq:measureTENO}
{\gamma _k} = {\left( C + \frac{{{\tau _6}}}{{{\beta _{k,r}} + \varepsilon }}\right)^q}{\text{ , }}k = 0,\cdots,3 ,
\end{equation}
where $\varepsilon  = {10^{ - 40}}$ is introduced to prevent the zero denominator. The parameters $C = 1$ and $q = 6$ are adopted for strong scale separation. $\beta _{k,r}$ can be defined similarly to WENO schemes \cite{Jiang1996}. 
A sixth-order $\tau _6$, which allows for good numerical stability with a reasonably large CFL number, can be constructed as \cite{fu2017targeted}
\begin{equation}
\label{eq:taukTENO}
\tau _6 = \left | {\beta _6} - \frac{1}{6}({\beta _{1,3}} + {\beta _{2,3}} + 4{\beta _{0,3}}) \right | = O(\Delta {x^6}),
\end{equation}
where the ${\beta _6}$ measures the global smoothness on the six-point full stencil.

\subsection{Weighting strategy of TENO6-opt}

For TENO6-opt scheme \cite{fu2016family}, the measured smoothness indicators are first normalized as
\begin{equation}
\label{eq:normalize}
{{\chi}_k } = \frac{{{\gamma _k}}}{{\sum\nolimits_{k = 0}^{3} {{\gamma _k}} }} ,
\end{equation}
and subsequently filtered by a sharp cut-off function 
\begin{equation}
\label{eq:cut-off}
{\delta _k} = \left\{ {\begin{array}{*{20}{c}}
0, &{\text{if }{\chi}_k  < {C_T},}\\
1, &{\text{otherwise},}
\end{array}} \right.
\end{equation}
where the parameter $C_T = 10^{-6}$ in this work can be determined by spectral analysis \cite{fu2016family}.

Finally, the weights $d_k$ subjected to the cut-off function $\delta _k$ are renormalized as
\begin{equation}
\label{eq:TENOweight_renormalize}
{w_k} = \frac{{d_k}{\delta _k}}{{\sum\nolimits_{k = 0}^{3} {d_k}{\delta _k} }}.
\end{equation}

In such a way, candidate stencils are identified to be either sufficiently smooth or nonsmooth sharply, and the contributions from candidate stencils containing discontinuities are fully removed. For smooth regions, all candidate stencils are judged to be smooth with $\delta _k = 1$ and thus the high-order accuracy of the background linear scheme is restored exactly without degeneration.

Note that, in order to get better resolution for small-scale flow structures, the weights $d_0 = 0.4294317718960898$, $d_1 = 0.1727270875843552$, $d_2 = 0.0855682281039113$, and $d_3 = 0.312272912415645$ are adopted to form the TENO6-opt scheme \cite{fu2017targeted}.

\section{A six-point NENO6 scheme}

The core idea of the TENO schemes is the binary selection of the candidate stencils based on the smoothness indicators with sufficient scale separation. Such a procedure can be interpreted as a nonlinear input-output mapping between the local flow field features and the binary selection of candidate stencils. By training a neural network with a properly constructed database, ANN can act as a black-box to mimic and even improve the TENO stencil selection procedure. To demonstrate this idea, we choose a simple deep neural network known as MLP to finish the procedure.   

\subsection{Neural network design} 
\label{sec:designNN}

As outlined in the introduction, we are interested in approximating a classifier of the form
\begin{equation}
\label{eq:appFunction}
{F} = {\mathbb{R}^{d_{in}}\rightarrow \mathbb{R}^{d_{out}}	},
\end{equation}
using ANN, 
and
\begin{equation}
\label{eq:appFunction}
 {\mathbb{R}^{d_{in}} = \mathbb{R}^{6},  \mathbb{R}^{d_{out}} = \mathbb{R}^{4} 	}.
\end{equation}
The objective of the mapping approximation is to predict the possibilities of the four candidate stencils to be selected for the final reconstruction based on the six cell values used for the six-point construction.

\begin{figure}[htbp]%
\centering   
  \includegraphics[width=1.0\textwidth]{ 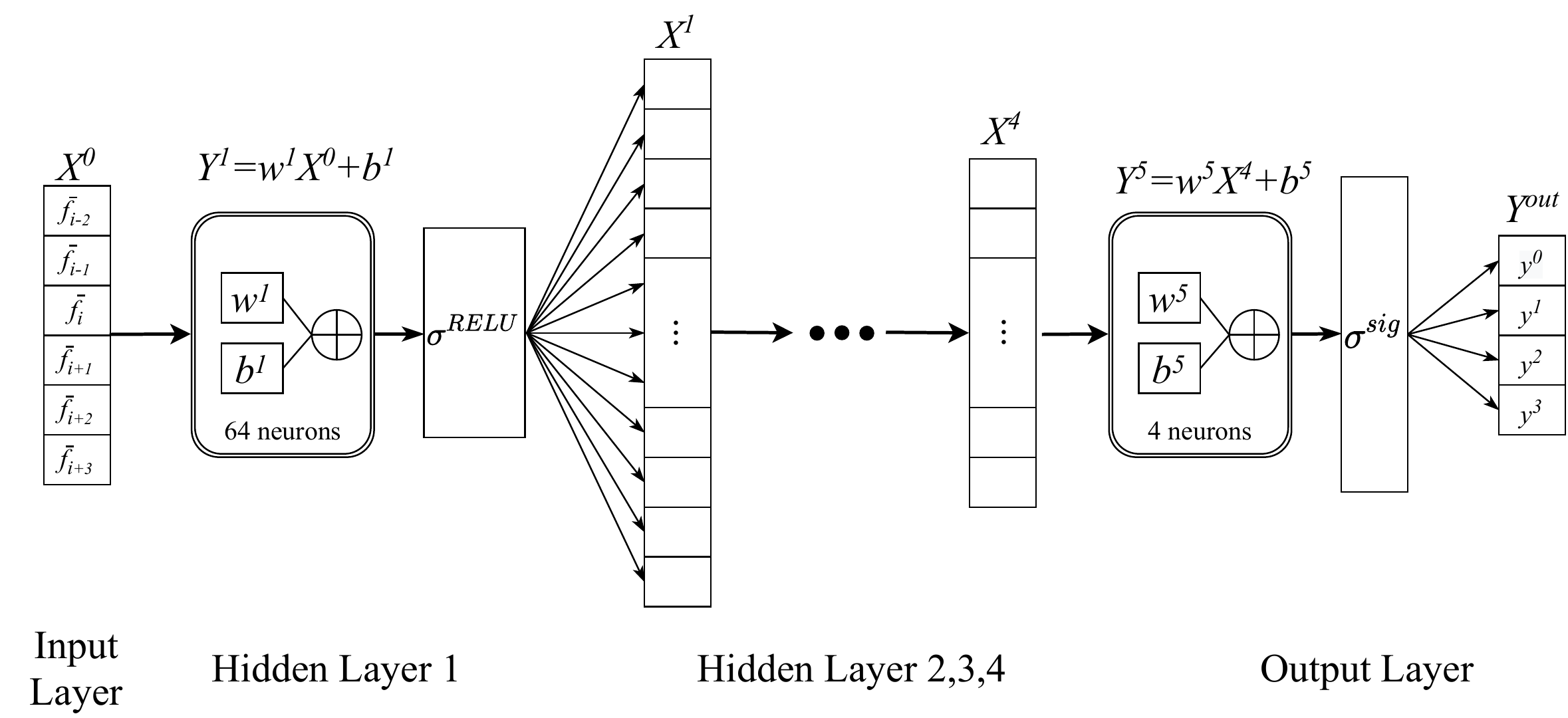}
\caption{Neural network architecture used for the candidate stencil selection in NENO6.}
\label{Fig:Network}
\end{figure}

An overview of the specific feed-forward network architecture known as MLP, in which the neurons are arranged in several layers, is provided in Fig.~\ref{Fig:Network}. The input layer provides an input signal to the network. While the last layer is the output layer, all the intermediate layers are denoted as the hidden layers. Four hidden layers, whose widths are 64, 32, 16 and 8 from the input layer to the output layer, respectively, are adopted in the present network architecture. 
The input for the neural network is the  normalized cell value vector $X^0$ =$[{\bar{f}}_{i-2},{\bar{f}}_{i-1},{\bar{f}}_{i},{\bar{f}}_{i+1},{\bar{f}}_{i+2},{\bar{f}}_{i+3}]^{\rm{T}}$ as given by
\begin{equation}
\label{eq:normalizainput}
{\bar{f}_{k}}=\frac{f_{k}}{\max(|f_{i-2}|,|f_{i-1}|,|f_{i}|,|f_{i+1}|,|f_{i+2}|,|f_{i+3}|,1)},      \text{  } k=i-2,i-1,\dotsi, i+3.
\end{equation}

Each hidden layer receives the output from the previous layer and performs an affine transformation
\begin{equation}
\label{eq:reconstuction}
{X^{l-1}\rightarrow Y^{l}} = w^{l}X^{l-1} + b^{l}, \text{ } l=1,2,3,4,5,
\end{equation}
where $w^l$ and $b^l$ are the weight matrix and bias vector at layer $l$, respectively. The output $Y^{l}$ is activated component-wise by a nonlinear activation function to form the input for the next layer. The activation function prevents the neural network from collapsing into a single affine transformation. Among many choices for activation functions, the rectified linear unit (RELU) activation function \cite{glorot2011deep}, i.e.
\begin{equation}
\label{eq:activationfunction}
{\sigma}^{\text{RELU}}(x) = { \text{max}(x,0)},\\
\end{equation}
is adopted as it mitigates the vanishing gradient problem and yields smooth output.
For the output layer made of four neurons, the vector $Y^{out}$ is obtained with a sigmoid function, i.e.
\begin{equation}
\label{eq:sigmoidfunction}
{\sigma}^{sig}{(x)} = \frac{1}{1+e^{-x}},\\
\end{equation}
instead of the activation function Eq.~(\ref{eq:activationfunction}) to project the output values in $Y^{out}$ between $[0,1]$.
Consequently, the output can be viewed as the probability that the candidate stencil is crossed by discontinues. Specifically, if the probability is larger than $50\%$, the candidate stencil will be discarded in the final reconstruction. 

\subsection{Generating the training and validation datasets}

For the training of an ANN with the ENO-like stencil selection property, the definition of a sufficiently broad spectrum of input sample $X^0$ and the corresponding labeled data indicating the desired output $\hat{Y}$ (referred to as the ground truth) is required.

The labelled ground-truth datasets are generated from both the analytic and non-analytic functions with three different mesh resolutions following Algorithm~\ref{alg:constructiontrainingsets}. The chosen mesh resolutions are $N = 50$, $100$, and $200$, and the proportion of the labeled data collected from each resolution is $40\%$, $40\%$ and $20\%$, respectively. The analytic functions are given in Table~\ref{Tab:dataset} of ~\ref{section:Apped} and their labels are given straightforwardly. The non-analytic functions are those evolved temporally through solving the linear advection or Burgers equations with the initial conditions listed in Table~\ref{Tab:dataset}. All temporal solutions $u(x,t)$ are analyzed by the optimal TENO6-opt weighting strategy and the ground-truth labeled data is obtained correspondingly, e.g., if the candidate stencil is judged as smooth and should be selected for the final reconstruction, it is labeled with $\eta = 0$.

\begin{algorithm} [t]
		\caption{Flowchart for the construction of the training and validation datasets}
		\label{alg:constructiontrainingsets}
		\begin{algorithmic}
		    \For {grid resolution $\textit{N} = [50, 100, 200]$}
		    \State 1. Choose a function $f(x) = u(x)$ in Table \ref{Tab:dataset} or its temporally evolved solution $f(x) = u(x, t)$ generated by solving the linear advection $\frac{\partial u}{\partial t} + \frac{\partial u}{\partial x} = 0$ or the Burgers function $\frac{\partial u}{\partial t} + \frac{\partial{( {u^2}/{2})}}{\partial x} = 0$; 			  
		    \Comment {\textit{See Table~\ref{Tab:dataset} for a detailed description of the function.}}
		    
		    \State 2. Pick the cell interface location $x_{i+1/2}$, and prepare the candidate stencils necessary for the stencil selection procedure;
		        \Comment {\textit{For the six-point reconstruction, see $\{S_0,S_1,S_2,S_3\}$ in Fig.\ref{Fig:incremental_stencil}.}}

		    \State 3. Prepare the input data, i.e. the normalized cell value vector $X^0$ =$[{\bar{f}}_{i-2},{\bar{f}}_{i-1},{\bar{f}}_{i},{\bar{f}}_{i+1},{\bar{f}}_{i+2},{\bar{f}}_{i+3}]^{\rm{T}}$ by applying Eq.~(\ref{eq:normalizainput});
		    
		    \State 4. Determine the labeled data, i.e. the ground-truth stencil selection output according to the regularity of the function $u(x)$;
		    \Comment{\textit{if one candidate stencil contains genuine discontinuities, then the corresponding labeled flag $\eta = 1$; otherwise, the labeled flag $\eta = 0$.}}
		 \EndFor
		\end{algorithmic}
\end{algorithm}

Through the aforementioned procedure, samples including smooth profiles, low-wavenumber profiles, high-wavenumber profiles, and genuine discontinuities are provided for the neuron network training and ultimately the training and validation sets of size 709000 are generated. At last, the first $70\%$ of the data is assigned to the training set while the remaining $30\%$ to the validation set.
\\

\subsection{Neural network training}

The cost function deployed to train the model is given by the binary cross entropy function
\begin{equation}
\label{eq:Crossentropy}
{\mathcal{L}_{BCE} = -\frac{1}{m} \sum_{m}{\left[\frac{1}{S} \sum_{k=1}^{S} {[ \hat{Y}_k^1\log(Y_k^1) + (1-\hat{Y}_k^1) \log(1-{Y}_k^1)]}\right]}_{m}},
\end{equation}
where $S$ denotes the number of neurons for the output layer, $m$ is the mini-batch size and $Y_k^1$, $\hat{Y}_k^1$ are the predicted and ground-truth non-smooth stencil probability distribution for a given sample respectively. The target of the optimization is to measure and minimize the discrepancy between the probability distribution of candidate stencil selection predicted by the MLP network and the ground-truth binomial distribution.
The training is performed using a stochastic optimization algorithm, which invokes mini-batches of size $m$ from the training dataset. The whole epoch $n_{epoch}$ is 1000 and the learning rate $L_r$ which begins with 0.0001 is reduced successively by a factor of $50\%$ at the epoch $[40,90,160,300,700]$.    
More specifically, the full training dataset is shuffled when one full epoch is completed and the process is repeated for several epochs. The shuffling operation introduces stochasticity in the training dataset, which ensures faster convergence. The parameters for the training process are summarized in Table~\ref{Tab:nnParameters}. 

\begin{table}[h]
  \centering
  \caption{Parameters for the network training.}
  \scriptsize
  \label{Tab:nnParameters}
  \newcommand{\tabincell}[2]{\begin{tabular}{@{}#1@{}}#2\end{tabular}}
  \begin{tabular}{>{\centering\arraybackslash}m{3.5cm}
                  >{\centering\arraybackslash}m{6.0cm}
                 }
  \toprule
   {Training optimizer}   &   Adam \cite{kingma2014adam}   \\
   {Mini-batch size}      &   $m = 175$   \\
   {Epochs}               &   $n_{epoch} = 1000$   \\
   {Learning rate}        &   $L_r = 0.0001$, reduced successively by a factor of $50\%$ at the epoch $[40,90,160,300,700]$   \\

  \bottomrule
  \end{tabular}
  \end{table}

\begin{figure}[htbp]%
\centering
  \includegraphics[width=0.65\textwidth]{ 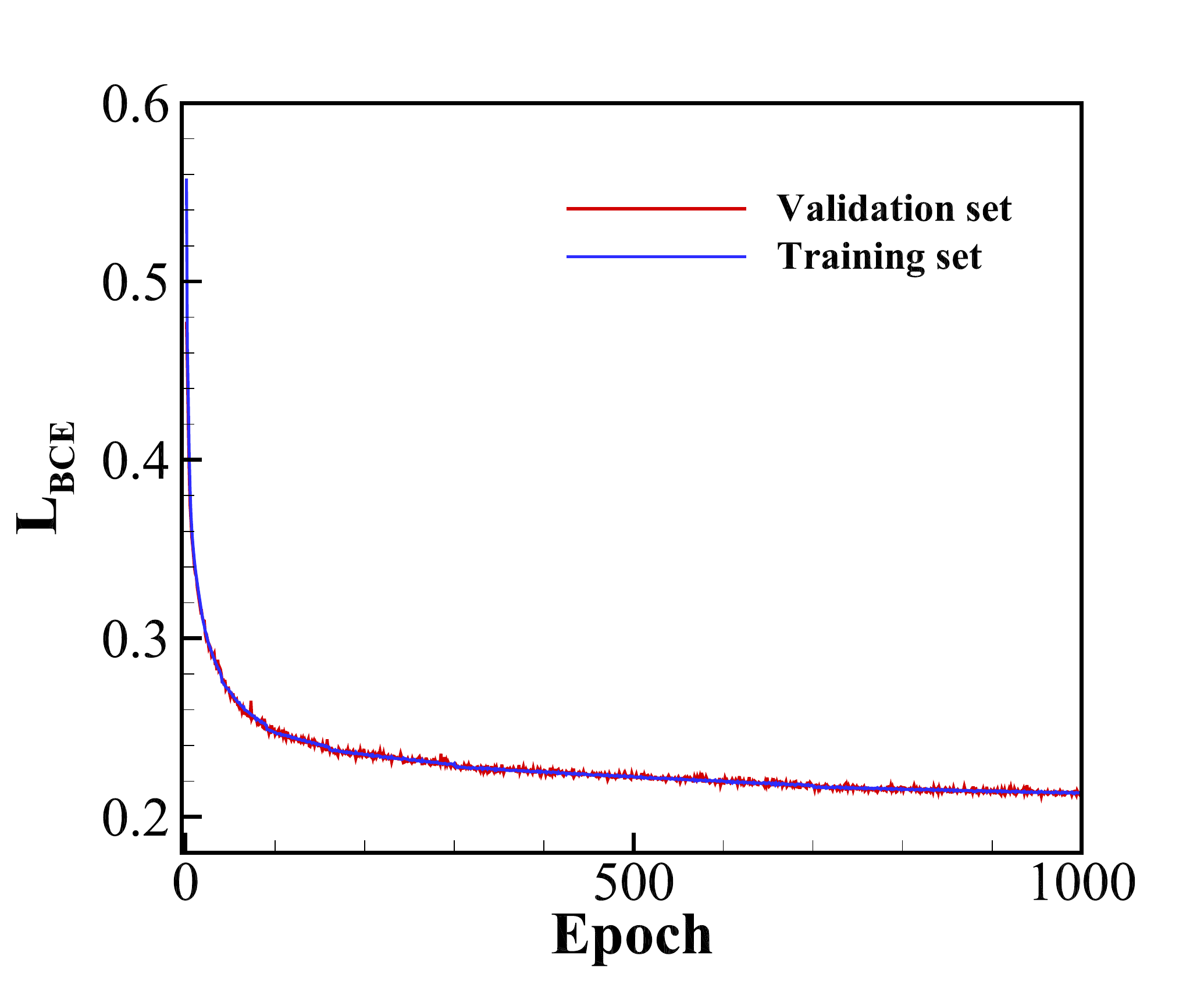}
\caption{Convergence history of the loss function $\mathcal{L}_{BCE}$ for the training and validation datasets.}
\label{Fig:epoch}
\end{figure}

Fig.~\ref{Fig:epoch} reports the convergence history of the loss function $\mathcal{L}_{BCE}$ throughout the optimization procedure with the samples belonging to both the training and validation datasets. It is observed that, with the number of epochs increasing, the loss function $\mathcal{L}_{BCE}$ decreases monotonically until convergence. Good convergence is also obtained for the validation dataset and no overfitting is observed.   

\subsection{Integration of the NENO6 scheme with a PDE solver}

For a neural network that has been trained offline based on the previously described dataset, each layer of the network can be fully characterized by the following parameters, i.e. the weight matrix $w^l$, the bias vector $b^l$, and the activation function $\sigma_{act}(\cdot)$. These parameters are then saved to interplay with an existing PDE solver following Algorithm~\ref{alg:IntegrationNENO_PDE}.

\begin{algorithm}
	\caption{Core algorithm for the integration of the ANN-based reconstruction scheme into a PDE solver}
	\label{alg:IntegrationNENO_PDE}
	\begin{algorithmic}
	
 	\State 1. Generation of features: provide the rescaled cell values $X^0$ =$[{\bar{f}}_{i-2},{\bar{f}}_{i-1},{\bar{f}}_{i},{\bar{f}}_{i+1},{\bar{f}}_{i+2},{\bar{f}}_{i+3}]^{\rm{T}}$ for six-point construction;
	
	\State 2. Predict the selection of candidate stencils: load the parameters $w^l$, $b^l$, and $\sigma_{act}(\cdot)$ into the PDE solver to determine the selection of candidate stencils;
	
 	\State 3. Calculate the high-order reconstructed flux at the cell interface $x_{i+1/2}$: based on the stencil selection results in step 2, re-normalize the weights and perform the convex combination to form the final reconstruction following Eq.~(\ref{eq:TENOweight_renormalize}) and Eq.~(\ref{eq:convex}), respectively;
	
	\State 4. Remedy for failure of stencil selection in step 2:
	
	  \If{all the candidate stencils are discarded in step 2 (rare event of failure of the ANN prediction)}
 	    \State Deploy the first-order upwind flux to replace the predicted flux in step 3; 
 	  \EndIf
 	\end{algorithmic}
\end{algorithm}

Based on the first four candidate stencils $\{S_0,S_1,S_2,S_3\}$ shown in Fig.~\ref{Fig:incremental_stencil}, cell values used for the six-point construction are rescaled and then fed into the trained network. According to the binary outputs, the ENO-like selection of candidate stencils is ensured. At last, the high-order reconstruction is formed by the convex combination of the candidate stencils based on the binary selection following Eq.~(\ref{eq:convex}). 

\section{Numerical validations}

In this section, a set of benchmark cases involving strong discontinuities and broadband flow scales is simulated. The NENO6 scheme is deployed to three different types of hyperbolic conservation laws to address its performance, including the linear advection equation, the Burgers equation and the Euler equation. Both one- and two-dimensional problems are considered. For multi-dimensional problems, the proposed reconstruction scheme is applied in a dimension-by-dimension manner based on the standard finite-difference framework. For comparison, the numerical results from the six-point WENO-CU6 \cite{hu2010adaptive} and TENO6-opt \cite{fu2016family}\cite{fu2018improved} schemes are provided. The optimal linear weights of the four candidate stencils for both TENO6-opt and NENO6 are the same as $d_0 = 0.4294317718960898$, $d_1 = 0.1727270875843552$, $d_2 = 0.0855682281039113$, and $d_3 = 0.312272912415645$. And the parameters used in WENO-CU6 are same as  \cite{hu2010adaptive}. 

\subsection{Accuracy order verification}
In this section, we consider a convergence study with a smooth function as
\begin{equation}
\label{eq:smoothorder}
{u(x)} = \cos(x).
\end{equation}

As shown in Table~\ref{Tab:Convergencesmooth},  the desired fourth-order accuracy is achieved for the nonlinear NENO6 scheme without degeneration. Note that the background linear weights of the NENO6/TENO6-opt scheme are optimized for better spectral properties with a fourth-order accuracy constraint \cite{fu2018improved}.

\begin{table}[h]
  \centering
  \caption{Convergence statistics of numerical error at $x = -0.5$ for Eq.~(\ref{eq:smoothorder}).}
  \scriptsize
  \label{Tab:Convergencesmooth}
  \newcommand{\tabincell}[2]{\begin{tabular}{@{}#1@{}}#2\end{tabular}}
  \begin{tabular}{>{\centering\arraybackslash}m{1.5cm}
                  >{\centering\arraybackslash}m{3.5cm}
                  >{\centering\arraybackslash}m{3.5cm}
                  >{\centering\arraybackslash}m{3.5cm}
                 }
  \toprule
  {N} & {Optimal linear scheme} &{TENO6-opt} &{NENO6}\\ 
  \hline

   {20}   &  1.0211e-06 (-)      & 1.0211e-06 (-)         & 1.0211e-06 (-)   \\
   {40}   & 6.37965e-08 (4.001)  & 6.37965e-08 (4.001)     & 6.37965e-08 (4.001)   \\
   {80}   & 3.98157e-09 (4.002)  & 3.98157e-09 (4.002)     & 3.98157e-09 (4.002)    \\
   {160}  & 2.48589e-10 (4.002)  & 2.48588e-10 (4.002)    & 2.48588e-10 (4.002)   \\
   {320}  & 1.5527e-11  (4.001)  & 1.55264e-11 (4.001)    & 1.55264e-11 (4.001)    \\
  \bottomrule
  \end{tabular}
    \begin{tablenotes}
        \item[a] Note that, for each table element, the error statistic is provided as well as the convergence order presented in the parentheses. 
    \end{tablenotes}  
\end{table}

\subsection{Linear advection}
In this section, we first consider the linear advection equation
\begin{equation}
\label{eq:linear_advect}
\frac{\partial u}{\partial t} + \frac{\partial u}{\partial x} = 0 ,
\end{equation}
with different initial and boundary conditions.

\subsubsection{Accuracy test}

To verify the convergence order of the proposed NENO6 scheme, the one-dimensional sine wave advection problem \cite{fu2016family} is considered. The linear advection equation with the initial condition
\begin{equation}
\label{eq:linear_advect_init}
{u}(x,0) = {\sin(\pi x)}, \text{ } 
\end{equation}
is solved in a computational domain $ 0 \leq x \leq 2$ and the final time is $t = 2$. Periodic boundary conditions are imposed at $ x = 0$ and $x = 2$.
\begin{table}[h]
  \centering
  \caption{Convergence statistics of numerical error with $L_\infty$ norm for the linear advection problem.}
  \scriptsize
  \label{Tab:sinorder}
  \newcommand{\tabincell}[2]{\begin{tabular}{@{}#1@{}}#2\end{tabular}}
  \begin{tabular}{>{\centering\arraybackslash}m{1.5cm}
                  >{\centering\arraybackslash}m{3.5cm}
                  >{\centering\arraybackslash}m{3.5cm}
                  >{\centering\arraybackslash}m{3.5cm}
                 }
  \toprule
  {N} & {Optimal linear scheme} &{TENO6-opt} &{NENO6}\\ 
  \hline

   {10}   &  0.0151104(-)        & 0.0151104(-)              & 0.0155788 (-)   \\
   {20}   &  0.00120265 (3.651)  & 0.00120265 (3.651)        & 0.00215545 (2.854)   \\
   {40}   &  7.96192e-05 (3.917)  & 7.96192e-05 (3.917)        & 8.77273e-05 (4.612)    \\
   {60}   &  1.58942e-05 (3.974)  & 1.58942e-05 (3.974)        & 1.76646e-05 (3.953)   \\
   {240}  &  6.31325e-08 (3.988)  &  6.31325e-08 (3.988)       & 1.18716e-06 (1.948)    \\
   {480}  &  3.91264e-09 (4.012)  & 3.91264e-09 (4.012)        & 1.00042e-07 (3.569)    \\
  \bottomrule
  \end{tabular}
    \begin{tablenotes}
        \item[a] Note that, for each table element, the error statistic with $L_\infty$ norm is provided as well as the convergence order presented in the parentheses. 
    \end{tablenotes}    
\end{table}

As shown in Table.~\ref{Tab:sinorder}, the desired fourth-order accuracy is achieved for the nonlinear TENO6-opt scheme without degeneration. For the NENO6 scheme, order degeneration occurs because of the accuracy degeneration of the neural network prediction. However, based on the present framework, the NENO6 scheme ensures at least a second-order convergence. In contrast, the data-driven WENO-NN scheme \cite{stevens2020enhancement} degenerates to first-order accuracy although the five-point stencil is employed (see Fig.~1 in \cite{stevens2020enhancement}).  

To further evaluate the performance of the trained networks, we examine the $F_1$ score \cite{fawcett2004roc} defined as
\begin{equation}
\label{eq:F1score}
F_{1} = {\frac{2}{\text{1/precision+1/recall}}}, \text{  and  }\\ 
 {\left\{{\begin{array}{*{40}{c}}
    \text{precision} = {\frac {\text{TP}}{\text{TP+FP}}},\\
    \text{recall} = {\frac{\text{TP}}{\text{TP+FN}}}, \end{array}} \right.}
\end{equation}
where TP, FP and FN denote the results of true positives, false positives and false negatives, respectively. The $F_1$ score is calculated from the precision and the recall of the test, where the precision is the number of correctly identified positive results (TP) divided by the number of all positive results (TP+FP), including those not identified correctly (FP), and the recall is the number of correctly identified positive results divided by the number of all samples (TP+FN) that should have been identified as positive. 

In this case, all samples should be labelled as smooth and TP denotes the number of correctly identified smooth samples. The results show that a $F_1$ score of $98.3\%$ can be achieved during the training.

\begin{table}[h]
  \centering
  \caption{Averaged F1 score statistics for the $\sin(\pi x)$ function during the advection with different resolutions.}
  \scriptsize
  \label{Tab:sinorder}
  \newcommand{\tabincell}[2]{\begin{tabular}{@{}#1@{}}#2\end{tabular}}
  \begin{tabular}{>{\centering\arraybackslash}m{1.5cm}
                  >{\centering\arraybackslash}m{3.5cm}
                  >{\centering\arraybackslash}m{3.5cm}
                  >{\centering\arraybackslash}m{3.5cm}
                 }
  \toprule
  {Resolution} & {TP} &{FN}  &{F1 score(\%)}\\ 
  \hline

   {40}   &  38  & 2     & 97.4    \\
   {60}   &  58  & 2     & 98.3   \\
   {150}  &  144  &  6   & 98.1    \\
   {480}  &  467  & 13   & 98.6    \\
  \bottomrule
  \end{tabular}
\end{table}
\subsubsection{Multiwave advection}

This case is taken from Jiang and Shu \cite{Jiang1996} and the initial condition is given as
\begin{equation}
\label{eq:threepart}
{u}(x,0) = \left\{ {\begin{array}{*{20}{c}}
{\frac{1}{6}[G(x-1, \beta, z-\theta)+ G(x-1, \beta, z+\theta) + 4G(x-1, \beta, z)],} &{\text{if }0.2 \le x < 0.4} ,\\
{1,}&{\text{else} \text{ if }0.6\le x \le 0.8} ,  \\
{1-|10(x-1.1)|,}&{\text{else} \text{ if }1.0\le x \le 1.2} ,  \\
{\frac{1}{6}[F(x-1, \alpha, a-\theta)+ F(x-1, \alpha, a+\theta) + 4F(x-1, \alpha, a)],} &{\text{else} \text{ if } 1.4 \le x < 
1.6} ,\\
{0,} &{\text{otherwise,}}
\end{array}} \right.
\end{equation}
where
\begin{equation}
\label{eq:intialsin}
G(x, \beta, z) = e^{-\beta(x-z)^2} , \\
F(x, \alpha, a) = \sqrt{\max(1-\alpha^2(x-a)^2, 0)} .
\end{equation}
The parameters in Eqs.~(\ref{eq:threepart}--\ref{eq:intialsin}) are given as
\begin{equation}
\label{eq:constant}
a = 0.5, \text{ } z = -0.7, \text{ } \theta =0.005, \text{ } \alpha =10, \text{ and } \beta = \frac{\log2}{36\theta^2}.
\end{equation}

The initial distribution consists of a Gaussian pulse, a square wave, a sharply peaked triangle and a half-ellipse arranged from left to right in the computational domain $x\in[0,2]$. The computation is performed with WENO-CU6, TENO6-opt and NENO6 on $N = 150$ uniformly distributed mesh cells and the final simulation time is $t = 6$.  

\begin{figure}[htbp]%
\centering
  \includegraphics[width=1.0\textwidth]{ 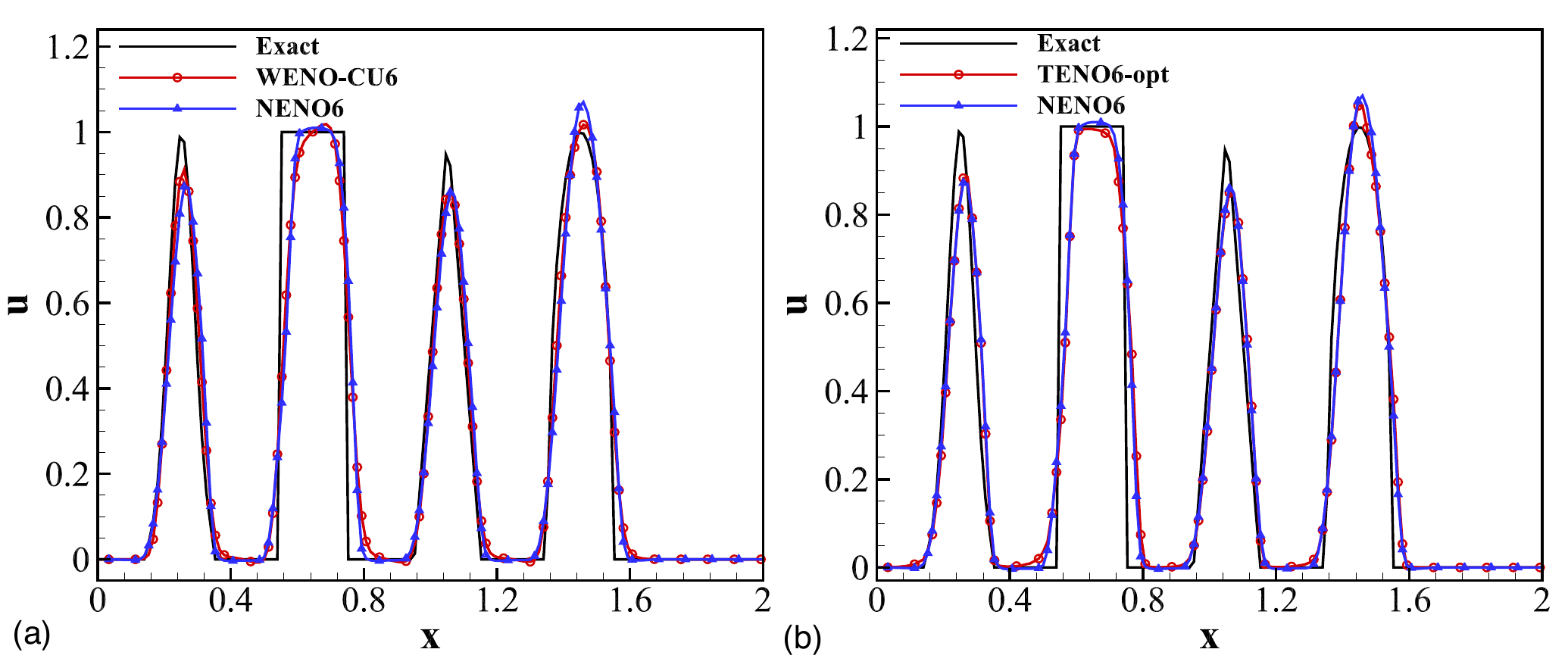}
\caption{Numerical results of advection of multi-wave with WENO-CU6, TENO6-opt and NENO6 schemes. Discretization is on 150 uniformly-distributed grid points. }
\label{Fig:full_multiwave}
\end{figure}

As shown in Fig.~\ref{Fig:full_multiwave}, for the advection of the half-ellipse wave, all schemes generate slight overshoots. Concerning the advection of the square wave, NENO6 resolves the discontinuities better than WENO-CU6 and TENO6-opt.

\begin{figure}[htbp]%
\centering
  \includegraphics[width=0.5\textwidth]{ 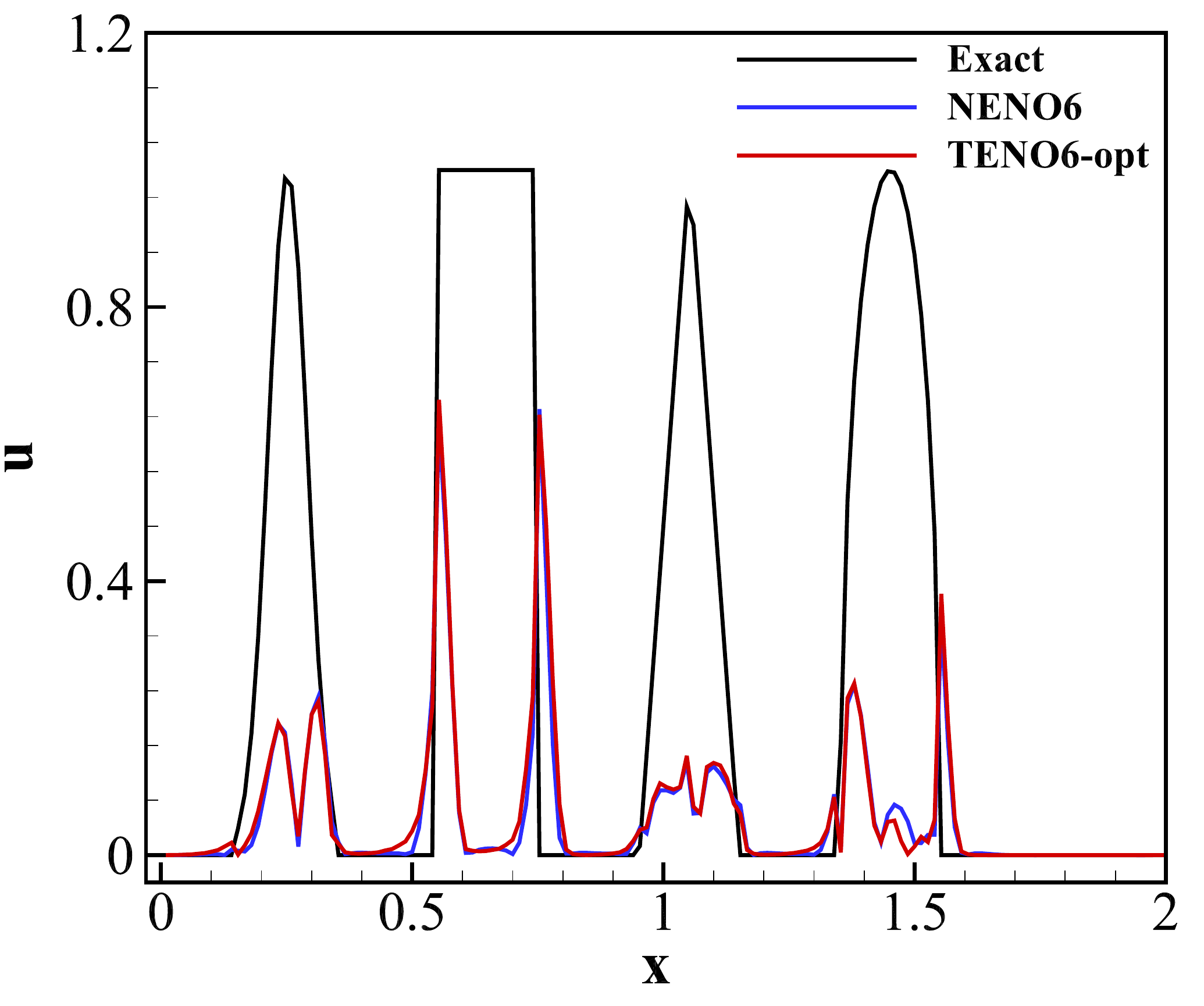}
\caption{Numerical error distributions from advection of multi-wave with TENO6-opt and NENO6 schemes at the final simulation time $t = 6$. The black line denotes the analytical solution while the blue and red lines denote the numerical errors from NENO6 and TENO6-opt, respectively. Discretization is on 150 uniformly-distributed grid points. }
\label{Fig:full_multiwave_error}
\end{figure}

To further assess the performance of the trained network quantitatively, as shown in Fig.~\ref{Fig:full_multiwave_error}, the numerical errors generated by TENO6-opt and NENO6 are examined at the final simulation time $t=6$. The results show that the NENO6 scheme captures the discontinuities better than TENO6-opt. 

\subsection{Burgers equation}
We consider the one-dimensional inviscid nonlinear Burgers equation \cite{sun2015slope}
\begin{equation}
\label{eq:Burgers}
\frac{\partial u}{\partial t} + \frac{\partial{( {u^2}/{2})}}{\partial x} = 0 .
\end{equation}
\subsubsection{Shockwave collision}
This case describes the collision of three shockwaves of varying strengths and speeds, which result in a single shock-wave moving from left to right. The initial condition is given by
\begin{equation}
\label{eq:shockcollision}
{u}(x,0) = \left\{ {\begin{array}{*{20}{c}}
{10,}&{\text{if } 0.0 \le x \le 0.2} ,\\
{6,}&{\text{else} \text{ if }  0.2 < x \le 0.4 } , \\
{0,}&{\text{else} \text{ if } 0.4 < x \le  0.6 },  \\
{-4,}&{\text{otherwise } 0.6 < x , } 
\end{array}} \right.
\end{equation}
on the domain [0,1] with open boundary conditions. The solution is marched in time until $t = 0.1$ with CFL = 0.1. 

\begin{figure}[htbp]%
\centering
  \includegraphics[width=1.0\textwidth]{ 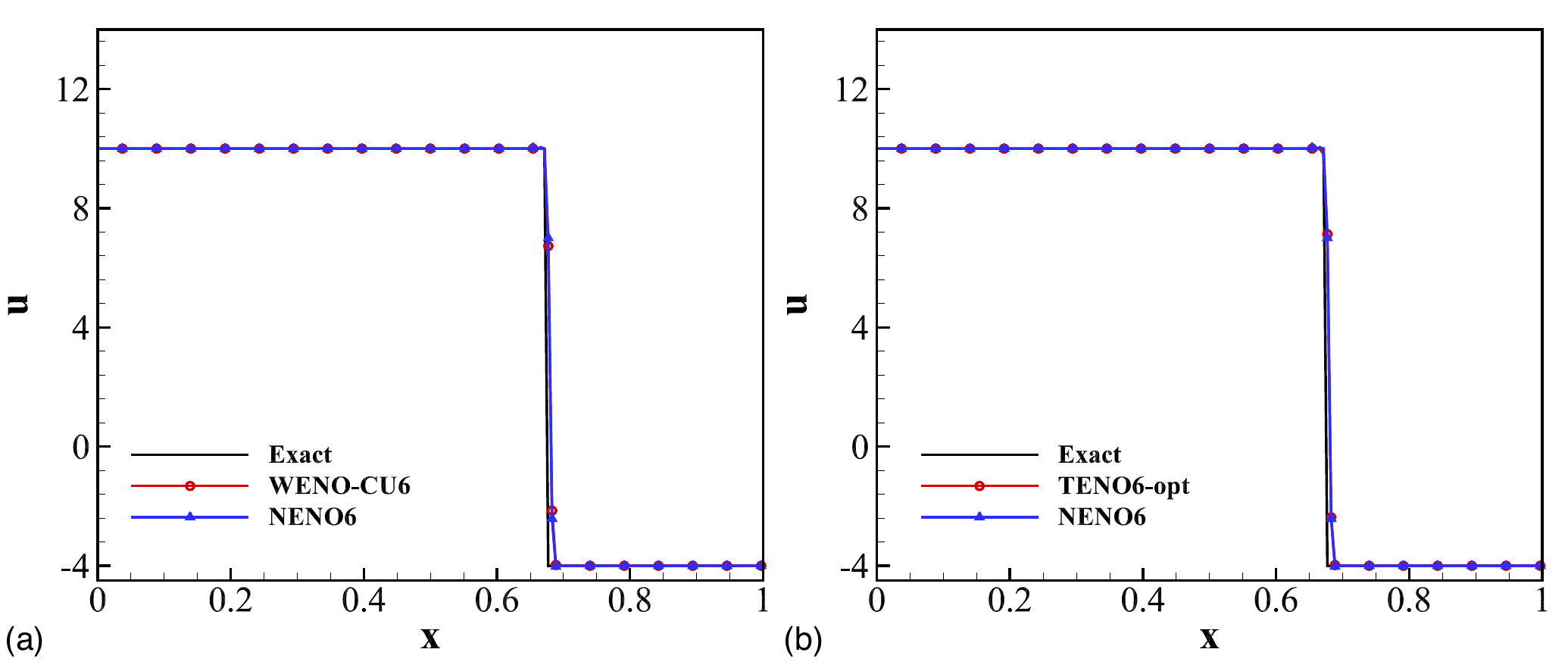}
\caption{Numerical results of the shock collision problem at time $t=0.1$ from WENO-CU6, TENO6-opt and NENO6 schemes. Discretization is on 175 uniformly distributed grid points. }
\label{Fig:collision}
\end{figure}

The results from WENO-CU6, TENO6-opt and NENO6 on a mesh with $N = 175$ cells are shown in Fig.~\ref{Fig:collision}. The NENO6 scheme performs as well as TENO6-opt and WENO-CU6 in capturing the moving shockwave.

\subsubsection{Compound wave}

We consider a case with complex compound waves, and the chosen initial condition consists of smooth and discontinuous profiles as
\begin{equation}
\label{eq:compound}
{u}(x,0) = \left\{ {\begin{array}{*{20}{c}}
{\text{sin}(\pi x),}&{\text{if }1 \le |x| \le 4} ,\\
{3,}&{\text{else} \text{ if } -1.0 < x \le -0.5 \text{  or } 0.0 < x \le 0.5 } , \\
{1,}&{\text{else} \text{ if } -0.5 < x \le  0.0 },  \\
{2,}&{\text{otherwise},} 
\end{array}}\right.
\end{equation}
on the domain [-4,4]. Periodic boundary conditions are imposed for the left and right sides. The solution is marched in time until $t = 0.4$ with CFL=0.2. The grid consists of 200 uniformly distributed cells.

\begin{figure}[htbp]%
\centering
  \includegraphics[width=1.0\textwidth]{ 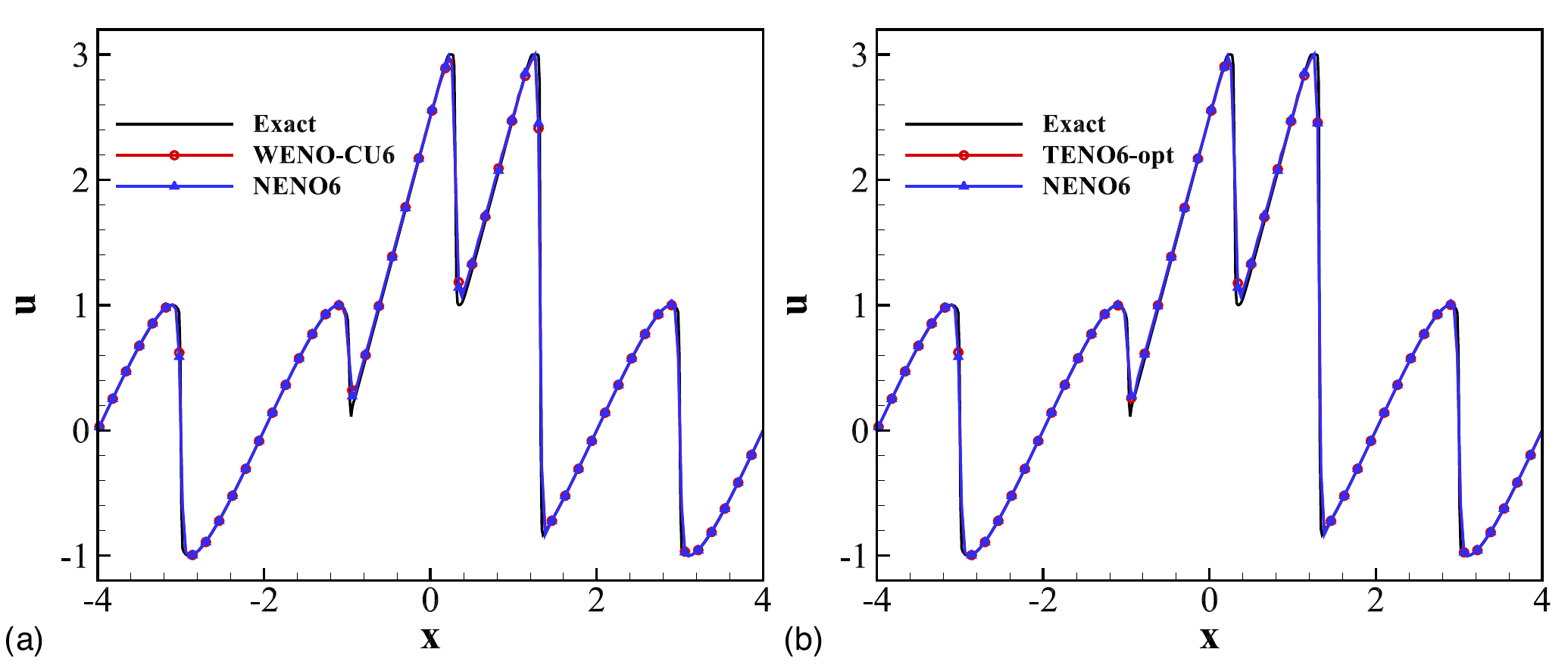}
\caption{Numerical results of the compound problem at time $t=0.4$ from WENO-CU6, TENO6-opt and NENO6 schemes. Discretization is on 200 uniformly distributed grid points. }
\label{Fig:compound}
\end{figure}

As shown in Fig.~\ref{Fig:compound}, WENO-CU6, TENO6-opt and NENO6 schemes provide similar results without any artificial numerical oscillation.

\subsection{Euler equations}

In this section, the NENO6 scheme is extended for systems of conservation laws. A set of test cases governed by the Euler equations is simulated. For multi-dimensional problems, the NENO6 scheme is applied in a dimension-by-dimension manner based on the standard finite-difference framework. First, the characteristic decomposition method based on the Roe average \cite{roe1981approximate} is employed and the Rusanov scheme \cite{Rusanov1961} is adopted for flux splitting. Afterwards, fluxes are treated as inputs of the MLP and the selections of sub-stencils used to compute interface flux are provided by the output of the MLP. The third-order strong-stability-preserving (SSP) Runge-Kutta method \cite{gottlieb2001strong} with a typical CFL number of 0.4 is adopted for the time advancement. 

\subsubsection{Lax's problem}

The Lax's problem \cite{Lax1954} is considered with the initial condition
\begin{equation}
\label{eq:lax}
(\rho ,u,p) = \left\{ {\begin{array}{*{20}{c}}
{(0.445,0.698,3.528),}&{\text{if }0 \le x < 0.5} ,\\
{(0.5,0,0.5710),}&{\text{if }0.5 \le x \le 1} ,
\end{array}} \right.
\end{equation}
and the final simulation time is $t = 0.14$.

\begin{figure}[htbp]%
\centering
  \includegraphics[width=1.0\textwidth]{ 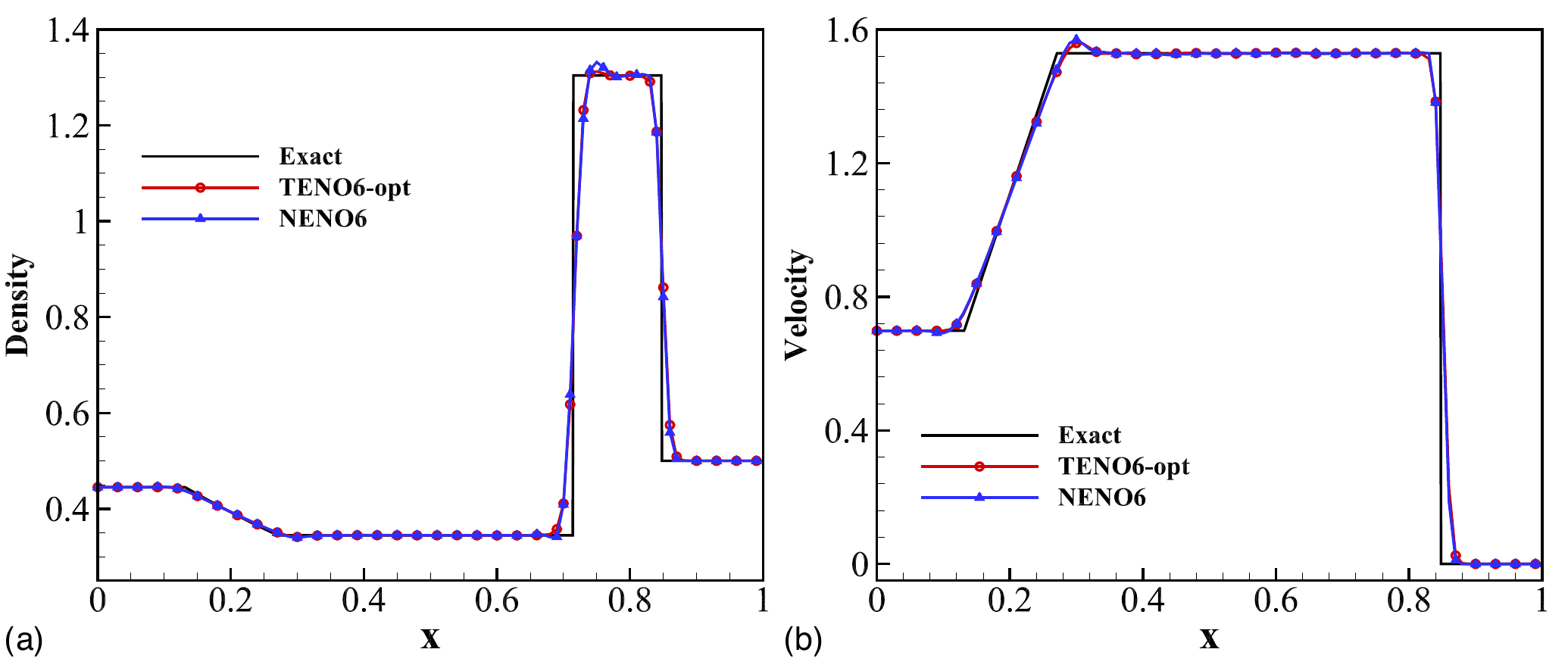}
\caption{Shock-tube problem: the Lax's problem. Density profile (a) and velocity profile (b) from TENO6-opt and NENO6 schemes. Discretization is on 100 uniformly distributed grid points and the final simulation time is $t = 0.14$.}
\label{Fig:lax}
\end{figure}

Fig.~\ref{Fig:lax} shows that both the contact and the shockwave are well captured with the NENO6 scheme. In the contact region, the NENO6 scheme generates a slightly larger overshoot than the TENO6-opt scheme. Similar behavior has been observed for other low-dissipation schemes, see Fig.~12 in \cite{fu2016family}.  
\subsubsection{Sod's problem}

The Sod's problem \cite{Sod1978} is considered with the initial condition
\begin{equation}
\label{eq:sod}
(\rho ,u,p) = \left\{ {\begin{array}{*{20}{c}}
{(1,0,1),}&{\text{if }0 \le x < 0.5} ,\\
{(0.125,0,0.1),}&{\text{if }0.5 \le x \le 1} ,
\end{array}} \right.
\end{equation}
and the final simulation time is $t = 0.2$. The computations are conducted on a mesh with 100 uniformly distributed grid points. 

\begin{figure}[htbp]%
\centering
  \includegraphics[width=1.0\textwidth]{ 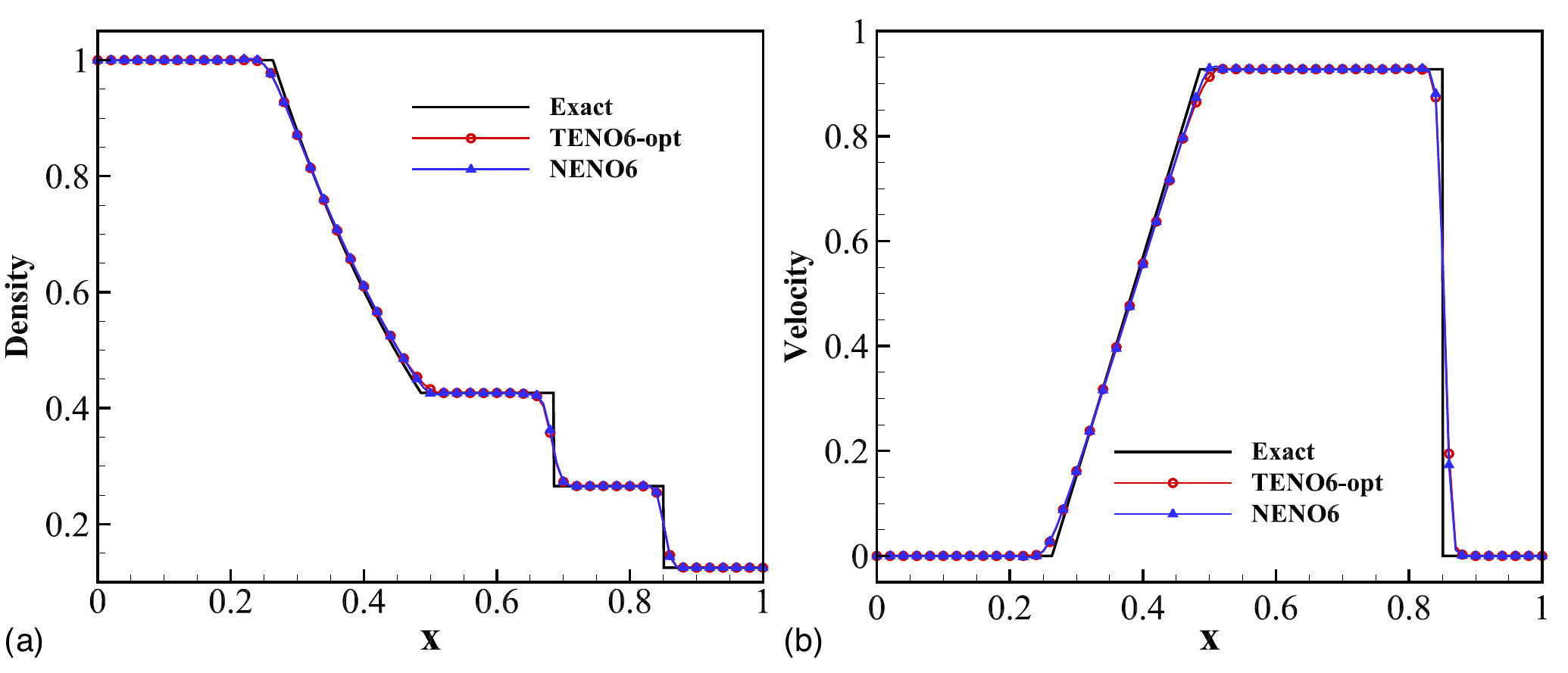}
\caption{Shock-tube problem: the Sod's problem. Density profile (a) and velocity profile (b) from TENO6-opt and NENO6 schemes. Discretization is on 100 uniformly distributed grid points and the final simulation time is $t = 0.2$. }
\label{Fig:sod}
\end{figure}

Fig.~\ref{Fig:sod} shows that the result from NENO6 is as good as that from TENO6-opt. No spurious oscillations can be observed and NENO6 performs slightly better in resolving the expansion fan.

\subsubsection{Shock density-wave interaction problem}
This case is taken from Shu and Osher \cite{Shu1989}. A one-dimensional Mach-3 shockwave interacts with a perturbed density field generating both small-scale structures and discontinuities. The initial condition is
\begin{equation}
\label{eq:shuosher}
(\rho ,u,p) = \left\{ {\begin{array}{*{20}{c}}
{(3.857,2.629,10.333),}&{\text{if } 0 \le x < 1} ,\\
{(1 + 0.2\sin (5(x-5)),0,1),}&{\text{else }  \text{if } 1 \le x \le 10} .
\end{array}} \right.
\end{equation}
The computational domain is [0,10] with $N = 200$ uniformly distributed mesh cells and the final evolution time is $t = 1.8$. The exact solution for reference is obtained by the fifth-order WENO-JS \cite{Jiang1996} scheme with $N = 2000$. 

\begin{figure}[htbp]%
\centering
  \includegraphics[width=1.0\textwidth]{ 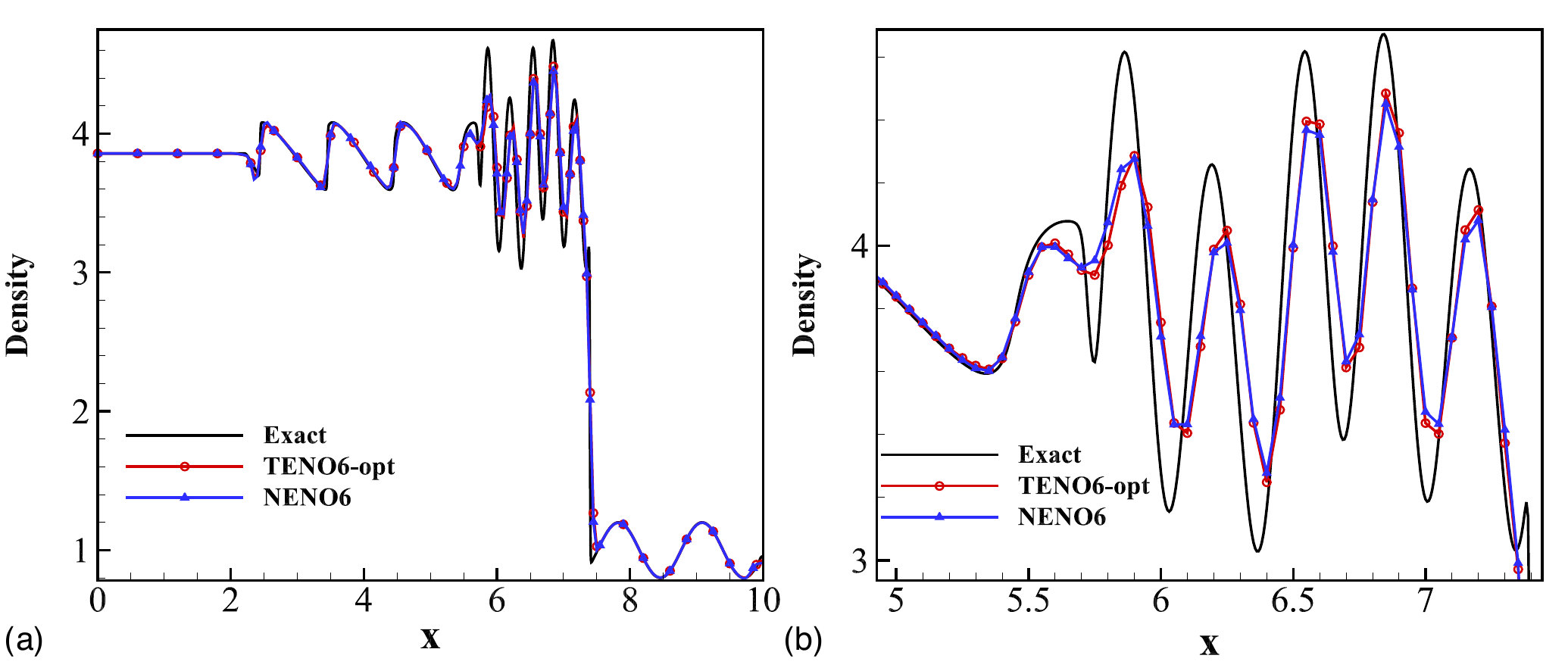}
\caption{Shock density-wave interaction problem: density distribution (a) and a zoomed-in view of the density distribution (b) from TENO6-opt and NENO6 schemes. Discretization is on 200 uniformly distributed grid points and the final simulation time is $t = 1.8$. }
 \label{Fig:shu-osher}
\end{figure}

Fig.~\ref{Fig:shu-osher} shows that the performance of the NENO6 scheme is similar to that of TENO6-opt. Both the shocklets and the high-wavenumber fluctuations are well resolved by NENO6. When compared to the data-driven WENO-NN scheme \cite{stevens2020enhancement}, the present result is free from the artificial overshoots with similar coarse resolutions, see their Fig.~12 in \cite{stevens2020enhancement}.

\subsubsection{Riemann problem: configuration 4}

Two-dimensional Riemann problems, first proposed in \citep{kurganov2002solution}, are classical benchmark cases for verifying numerical methods by solving the Euler equations. The 2D Riemann problem of configuration 4 is considered. The computational domain is $[0, 1] \times [0,1]$ and the final simulation time is $t = 0.25$. The initial condition is given as
\begin{equation}
\label{eq:Riemann2D04}
(\rho,u,v,p) = \left\{ {\begin{array}{*{20}{c}}
{(0.5065,0.8939,0,0.35),}&{\text{if }0.0 < x < 0.5, 0.5 < y < 1.0},\\
{(1.1,0,0,1.1),}&{\text{else } \text{if }0.5 < x < 1.0, 0.5 < y < 1.0},\\
{(1.1,0.8939,0.8939,1.1),}&{\text{else } \text{if }0.0 < x <0.5, 0.0 <y < 0.5}, \\
{(0.5065,0,0.8939,0.35),}&{\text{else } \text{if }0.5 < x <1.0, 0.0 <y< 0.5}. \\
\end{array}} \right.
\end{equation}

\begin{figure}[htbp]
\centering
\includegraphics[width=1.0\textwidth]{ 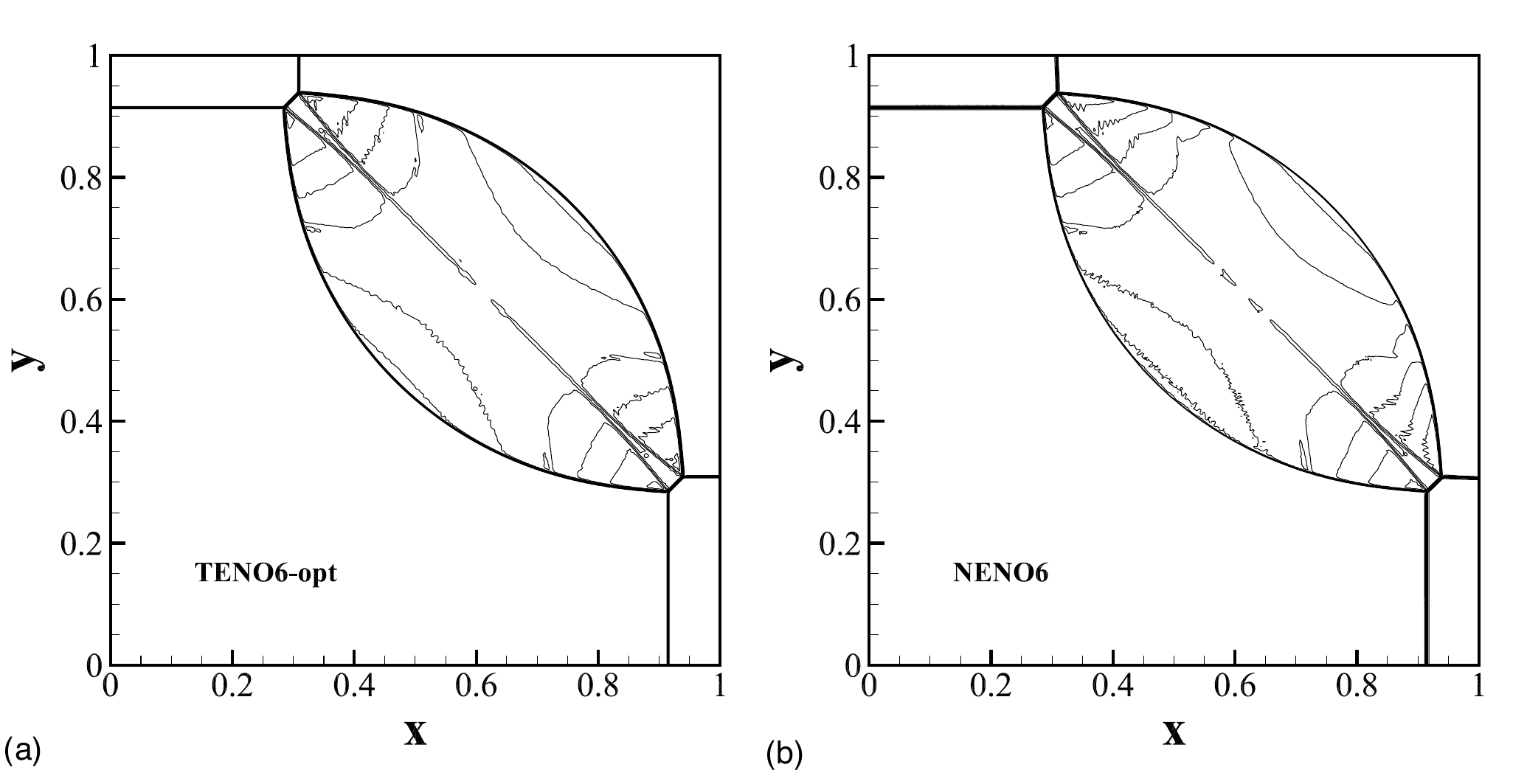}
  \caption{2D Riemann problem of configuration 4: density contours from TENO6-opt (a) and NENO6 (b) with the resolution of $480 \times 480$ at the simulation time $t = 0.25$. This figure is drawn with 22 density contours between 0.255 and 1.9.}
 \label{Fig:Riemann4}
\end{figure}

As shown in Fig.~\ref{Fig:Riemann4}, NENO6 captures the shock-wave patterns as sharp as TENO6-opt and the overall flow structures are well resolved by NENO6. It is worth noting that, without any parameter tuning, the good performance of the NENO6 scheme in the present 2D simulation demonstrates the generality of the proposed ANN-based high-order scheme. As a comparison, the previous ANN-based schemes are not straightforward to be extended for high-dimension simulations \cite{stevens2020enhancement}\cite{bar2019learning}.

\subsubsection{Riemann problem: configuration 6}

The 2D Riemann problem of configuration 6 is considered. The computational domain is $[0, 1] \times [0,1]$ and the final simulation time is $t = 0.3$. The initial condition is given as
\begin{equation}
\label{eq:Riemann2D06}
(\rho,u,v,p) = \left\{ {\begin{array}{*{20}{c}}
{(2.0,0.75,0.5,1.0),}&{\text{if }0.0 < x < 0.5, 0.5 < y < 1.0},\\
{(1.0,0.75,-0.5,1.0),}&{\text{else } \text{if }0.5 < x < 1.0, 0.5 < y < 1.0},\\
{(1.0,-0.75,0.5,1.0),}&{\text{else } \text{if }0.0 < x <0.5, 0.0 <y < 0.5}, \\
{(3.0,-0.75,-0.5,1.0),}&{\text{else } \text{if }0.5 < x <1.0, 0.0 <y< 0.5}. \\
\end{array}} \right.
\end{equation}

\begin{figure}[htbp]
\centering
\includegraphics[width=1.0\textwidth]{ 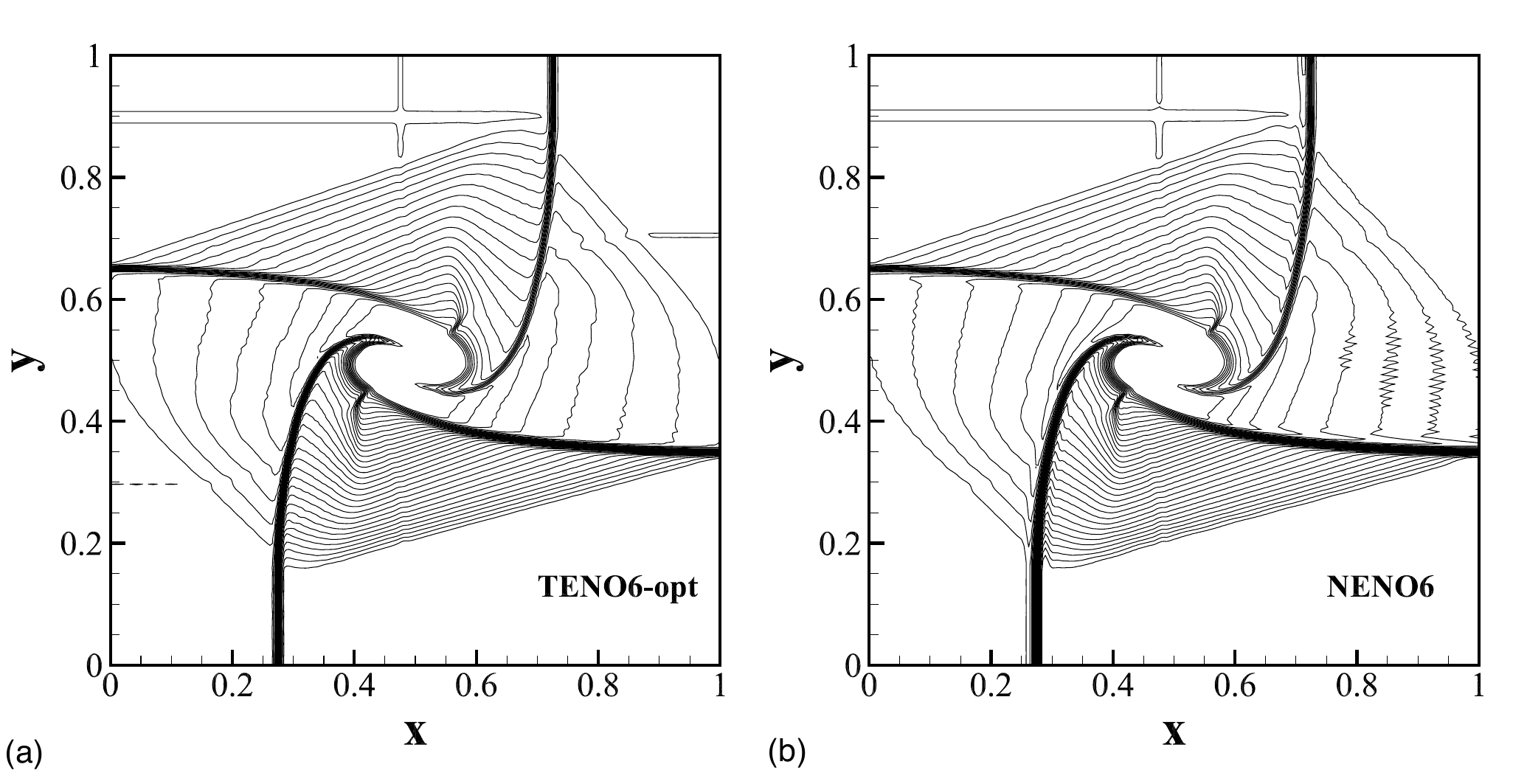}
  \caption{2D Riemann problem of configuration 6: density contours from TENO6-opt (a) and NENO6 (b) with the resolution of $480 \times 480$ at the simulation time $t = 0.3$. This figure is drawn with 36 density contours between 0.46 and 2.931.}
 \label{Fig:Riemann6}
\end{figure}

As shown in Fig.~\ref{Fig:Riemann6}, the performance of NENO6 is as good as TENO6-opt in terms of sharp contact discontinuities interface capturing.

\subsubsection{Double Mach reflection of a strong shock}

This two-dimensional case is taken from Woodward and Colella \cite{Woodward1984} with the initial condition as
\begin{equation}
\label{eq:DMR}
(\rho ,u,v,p) = \left\{ {\begin{array}{*{20}{c}}
{(1.4,0,0,1) ,}&{\text{if } y < 1.732(x - 0.1667) ,}\\
{(8,7.145, - 4.125,116.8333) , }&{\text{otherwise}.}
\end{array}} \right.
\end{equation}
The computational domain is $[0,4] \times [0,1]$ and the final simulation time is $t = 0.2$. 

\begin{figure}[htbp]
\centering
\includegraphics[width=1.0\textwidth]{ 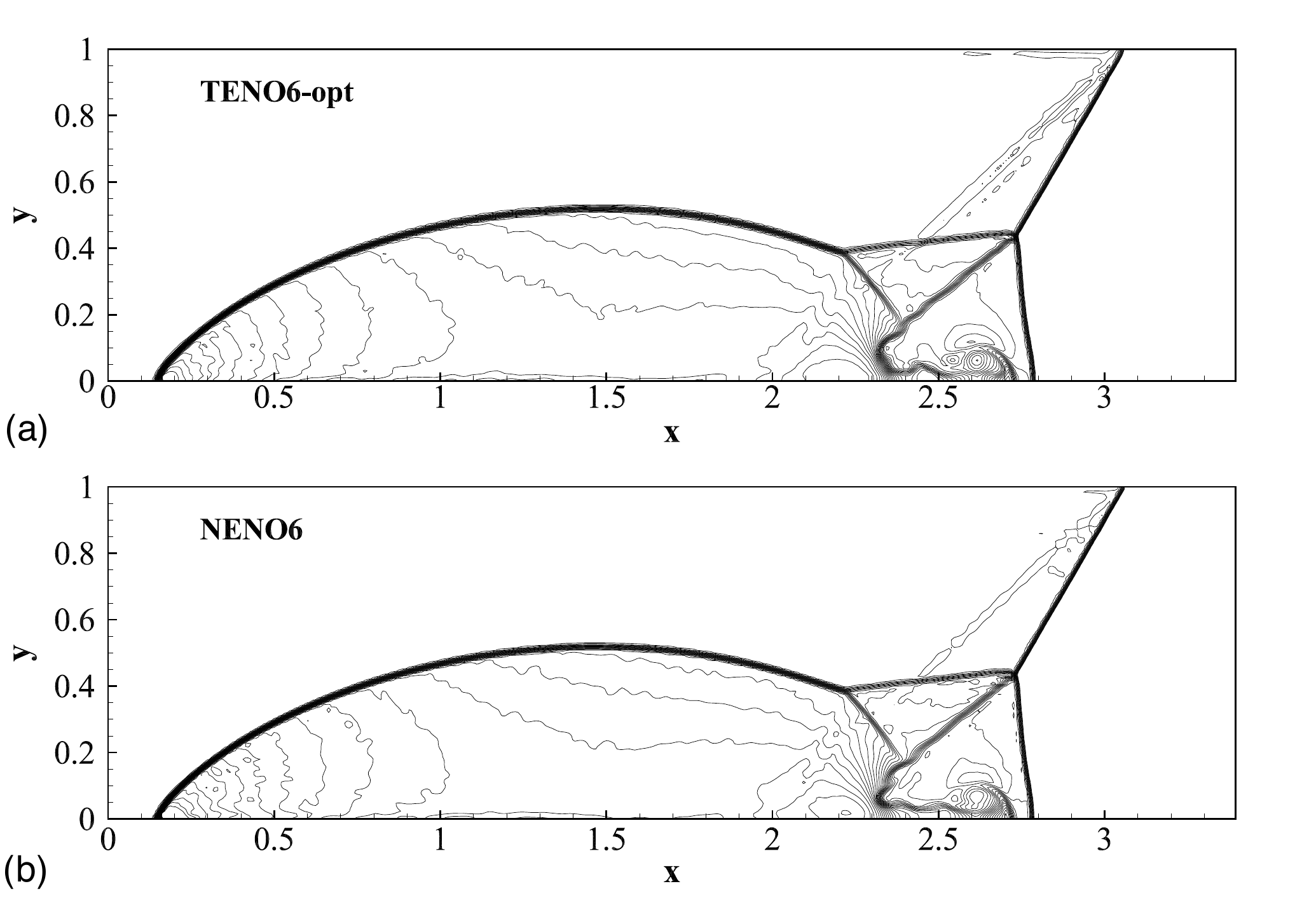}
  \caption{Double Mach reflection of a strong shock: density contours from TENO6-opt and NENO6 schemes at the simulation time $t = 0.2$. Resolution is $512 \times 128$. This figure is drawn with 42 density contours between 3.27335 and 20.1335.}
 \label{Fig:dmr512}
\end{figure}

\begin{figure}[htbp]
\centering
\includegraphics[width=1.0\textwidth]{ 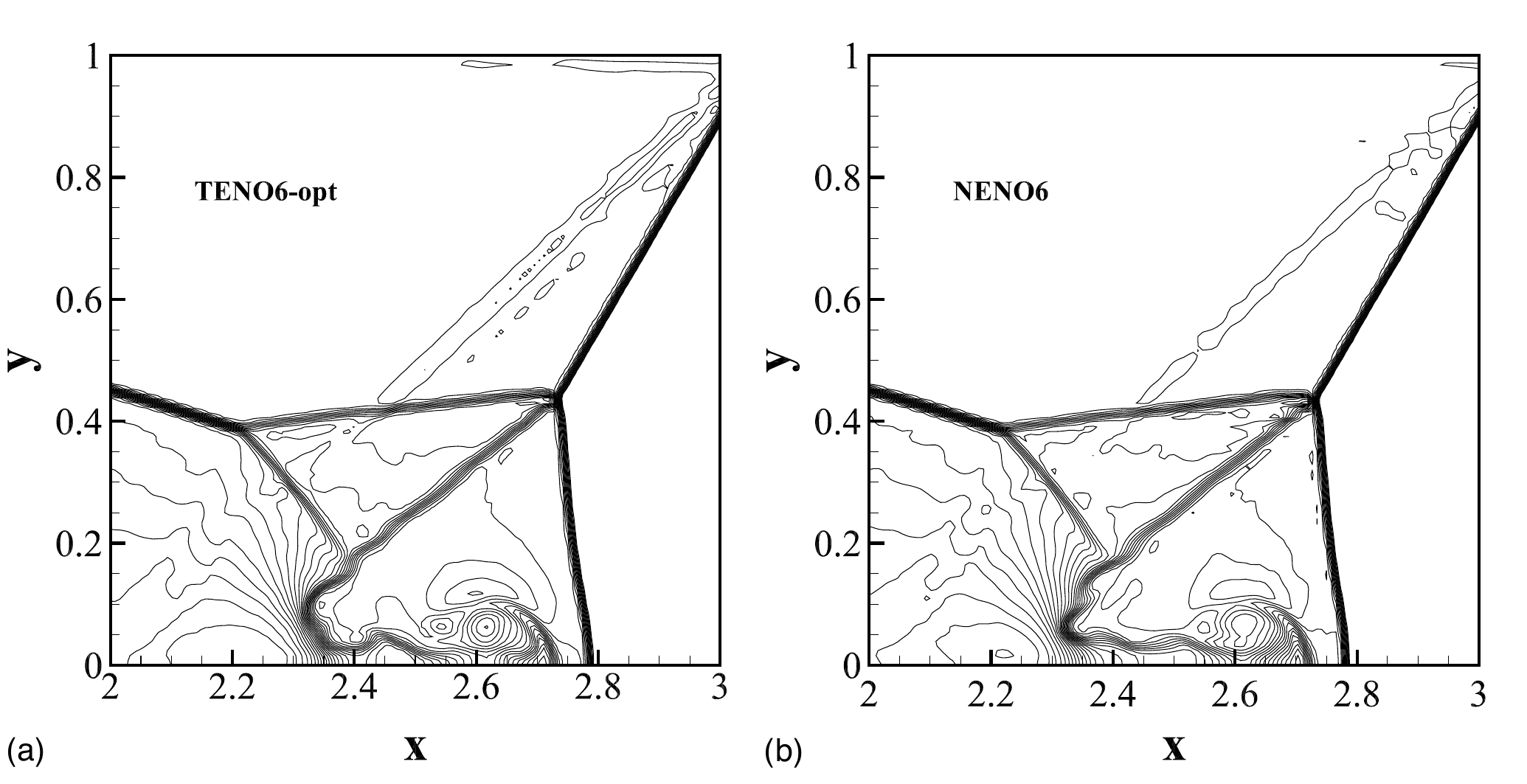}
  \caption{Double Mach reflection of a strong shock: a zoomed-in view of density contours from TENO6-opt and NENO6 schemes at the simulation time $t = 0.2$. Resolution is $512 \times 128$. This figure is drawn with 42 density contours between 3.27335 and 20.1335.}
 \label{Fig:dmr512zoomin}
\end{figure}

\begin{figure}[htbp]
\centering
\includegraphics[width=1.0\textwidth]{ 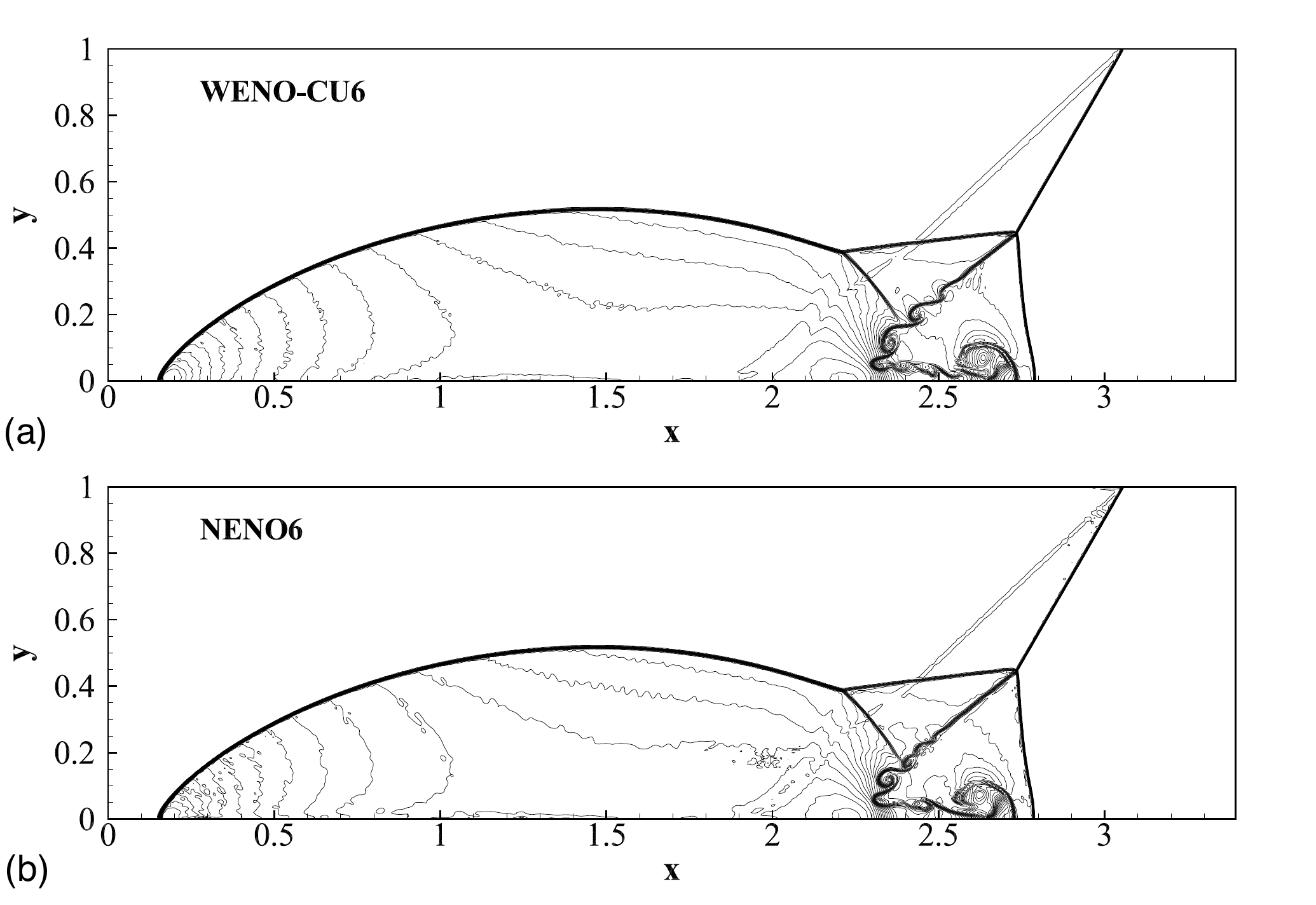}
  \caption{Double Mach reflection of a strong shock: density contours from WENO-CU6 and NENO6 schemes at the simulation time $t = 0.2$. Resolution is $1024 \times 256$. This figure is drawn with 42 density contours between 3.27335 and 20.1335.}
 \label{Fig:dmr1024H}
\end{figure}

\begin{figure}[htbp]
\centering
\includegraphics[width=1.0\textwidth]{ 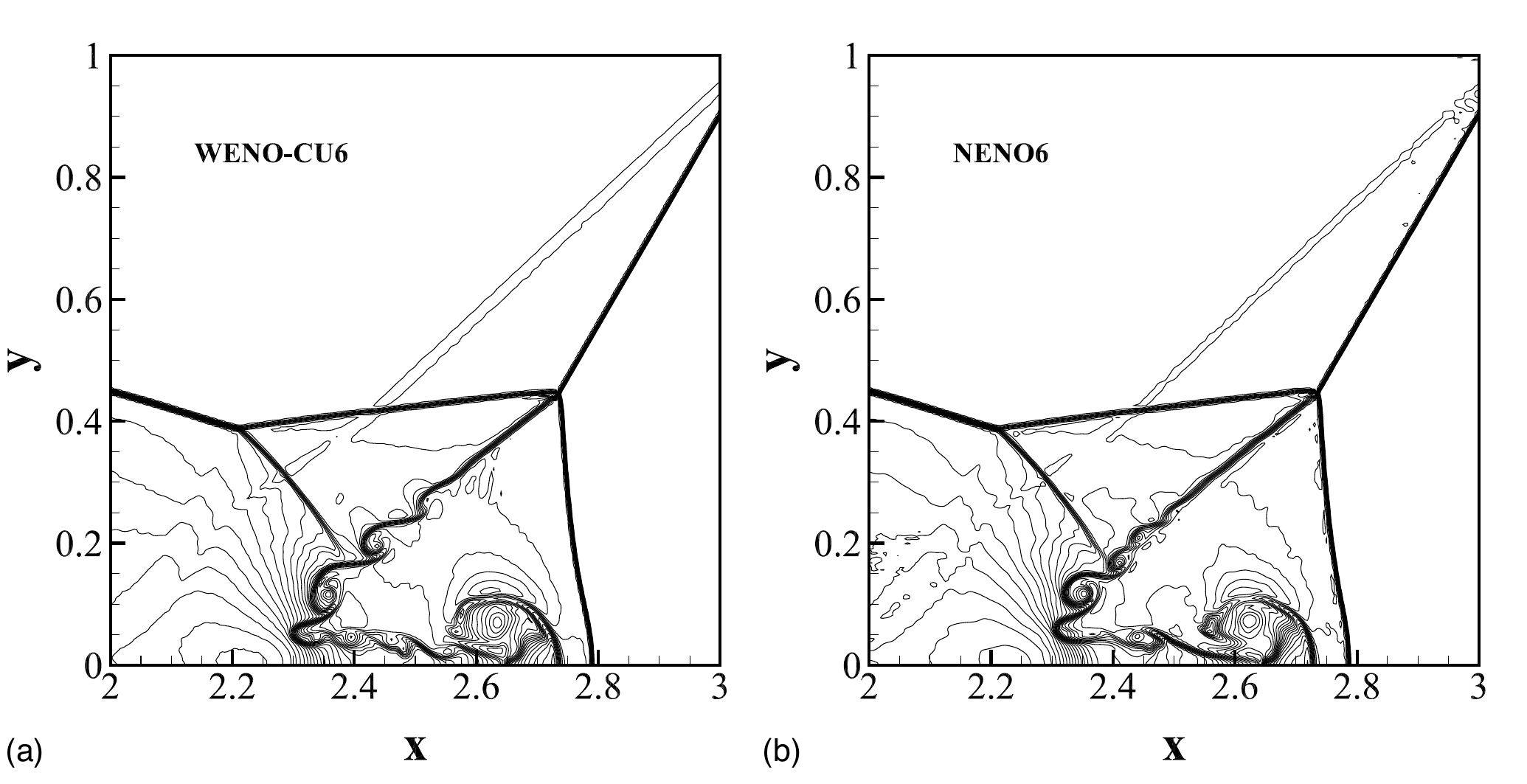}
  \caption{Double Mach reflection of a strong shock: a zoomed-in view of density contours from WENO-CU6 and NENO6 schemes at the simulation time $t = 0.2$. Resolution is $1024 \times 256$. This figure is drawn with 42 density contours between 3.27335 and 20.1335.}
 \label{Fig:dmr1024zoominH}
\end{figure}

The computed density distributions from TENO6-opt and NENO6 with resolution of $512 \times 128$ are given by Fig.~\ref{Fig:dmr512} and Fig.~\ref{Fig:dmr512zoomin}. The proposed NENO6 scheme shows good performance in capturing the complex shockwaves as well as in resolving the instabilities in the blow-up regions.
The higher resolution results from WENO-CU6 and NENO6 are given in Fig.~\ref{Fig:dmr1024H} and Fig.~\ref{Fig:dmr1024zoominH}. With the resolution of $1024 \times 256$, the NENO6 scheme is robust and shows an improved resolution for shock-capturing when compared to WENO-CU6. 

Compared with other ANN-based high-order ENO-type schemes \cite{bar2019learning}\cite{stevens2020enhancement}, for which the applications are restricted to one-dimensional prototype problems, NENO6 exhibits excellent robustness, low dissipation and shock-capturing capability in complicated multi-dimensional problems. 

\section{Conclusions}
In this paper, a six-point data-driven high-order shock-capturing NENO6 scheme is proposed and its performance is demonstrated by conducting a set of critical benchmark cases. The conclusions are as follows.
\begin{itemize}
    \item The new framework establishes a novel ENO-like paradigm for shock-capturing based on ANN. It rephrases the empirical smoothness measures of TENO6-opt scheme by a data-driven method. There are basically two steps towards building high-order NENO6 scheme: (i) train and deploy a neural network as a black box to achieve the ENO-like stencil selection procedure; (ii) assemble the high-order reconstruction by combining the candidate fluxes with optimal linear weights.
    
    \item Based on the ENO-like stencil selection indicated by ANN, the resultant scheme maintains boundedness of the reconstruction by the convex combination of the low-order polynomial-based candidate fluxes. This convex combination also guarantees that the contribution of candidate flux crossed by discontinuities vanishes completely while the high-order accuracy is restored in smooth regions. By generating a broad variety of input data, low dissipation and shock-capturing properties of the assembled NENO6 scheme can be simultaneously achieved. 
    
    \item Without further case-by-case tuning, the present framework applies to 1D and 2D problems governed by different hyperbolic conservation laws, such as the linear advection problems, the Burgers problems and the Euler problems. Similarly to the classical high-order schemes, the NENO6 scheme can be deployed straightforwardly to practical applications, suggesting an improved generality when compared to previous ANN-based high-order schemes. 

    \item A set of critical benchmark cases is simulated. Numerical results demonstrate the good capability of the NENO6 scheme in terms of recovering the desired high-order accuracy in smooth regions (at least second-order accuracy can be guaranteed if the prediction of ANN does not fail), preserving low numerical dissipation for resolution of physical fluctuations, and sharp capturing of discontinuities. 
    
    \item The present work reveals the potential of exploiting neural network to design high-order numerical method for hyperbolic conservation laws. Although the ANN is capable of essentially mimicking or even improving the state-of-the-art high-order ENO-type schemes, as a common issue in other ANN-based methods \cite{abgrall2020neural}, the quantity of the invoked neurons and the associated cost of the network training and evaluation are substantial, and further efficiency improvement is needed.

\end{itemize}
Considering the flexibility and the robustness of the present data-driven framework, future work will further improve the performance of the NENO6 scheme, and optimize the spectral properties for reliable implicit large-eddy simulations.

\section*{Declaration of Competing Interest}

The authors declare that they have no known competing financial interests or personal relationships that could have appeared to influence the work reported in this paper.

\section*{Acknowledgements}

The first author is partially supported by China Scholarship Council (No. 201706290041). The first author wishes to acknowledge Pähler Ludger for helpful discussions on training the present neural network. Lin Fu acknowledges the fund from Guangdong Basic and Applied Basic Research Foundation (No. 2022A1515011779), the fund from Shenzhen Municipal Central Government Guides Local Science and Technology Development Special Funds Funded Projects (No. 2021Szvup138), and the fund from Key Laboratory of Computational Aerodynamics, AVIC Aerodynamics Research Institute.

\appendix
\section{Functions for generating the training datasets}
\label{section:Apped}

Detailed descriptions of the functions and their built-in parameters for generating the training and validation datasets are as follows:

\begin{table}[h]
  \centering
  \caption{Parameters for training and validation datasets generation.}
  \scriptsize
  \label{Tab:dataset}
  \newcommand{\tabincell}[2]{\begin{tabular}{@{}#1@{}}#2\end{tabular}}
  \begin{tabular}{>{\centering\arraybackslash}m{0.05cm}
                  >{\centering\arraybackslash}m{3.3cm}
                  >{\centering\arraybackslash}m{1.0cm}
                  >{\centering\arraybackslash}m{4.0cm}
                  >{\centering\arraybackslash}m{1.5cm}
                  >{\centering\arraybackslash}m{1.8cm}
                  }
  \toprule
 No. & $u(x)$ & \text{Domain} & \text{Additional parameters} & \text{Resolution} & \text{Flag $\eta=(1-\delta)$}\\ 
  \hline

 1  & {$kx$}            & [-1 , 1] &  $k\in$ $\mathbb{U}$[-5, 5]                 &  $[50, 100, 200]$ & 0   \\

 2  &   {$a_k \sin(kx)$}            & [-1 , 1] &  $k\in$ $\mathbb{U}$[1, 5]             &  $[50, 100, 200]$ & 0   \\

 3  &  {$e^{a_1(x-a_2)^2}$}  & [-1 , 1] & $a_1$, $a_2$ $\in$ $\mathbb{U}$[-1, 1]  &  $[50, 100, 200]$  & 0  \\

 4  &   {$(r/P)(P-|(x $ mod $ (2P))-P|)$}  & [-1 , 1] & $r$ $\in$ $\mathbb{U}$[1, 3], $P$ $\in$ $\mathbb{U}$[$2\Delta x$, $8\Delta x$]  &  $[50, 100, 200]$  & 0/1  \\

 5  &  {$\max(0,b_1(x-x_0))+c$}  & [-1 , 1] &  $b_1$ $\in$ $\mathbb{N}$[0, 0.6], $x_0$ $\in$ $\mathbb{U}$[-0.6,0.6], $c$ $\in$ $\mathbb{U}$[0, 1] &  $[50, 100, 200]$  & 0/1 \\

 6  &  {$(x-a_1)^{k_1}+(x-a_2)^{k_2}$}  & [-1 , 1] & $k_i$ $\in$ $\mathbb{U}$[2,8], $a_1, a_2$ $\in$ $\mathbb{U}$[0, 2]  &  $[50, 100, 200]$  & 0/1 \\
   
 7  &  $\sqrt{\max(0,1-\alpha^2(x-a)^2)}$ & [-1 , 1] &  $\alpha$, a $\in$ $\mathbb{U}$[0, 2]  &  $[50, 100, 200]$ & 0/1   \\
   
 8  & ${\left\{ {\begin{array}{*{20}{c}}
   {x_l,} & {\text{if } x \le x_0 } ,\\
   {x_r,} & {\text{otherwise}}. \end{array}} \right.}$ & $[-1,1]$ &  $x_l,x_r, x_0$ $\in$ $\mathbb{U}$[0,2]  &  $[50, 100, 200]$ & 1   \\
   
 9  & $a_1 \sin(N_f \pi (x-x_0))$  & [-1,1] &  $a_1, x_0 $ $\in$ $\mathbb{U}$[0, 2]  &  $[50, 100, 200]$ & 0/1   \\
  \bottomrule 
  \end{tabular}
    \begin{tablenotes}
        \item[a] Note that $\mathbb{U}$[a,b] denotes the uniform distribution on the interval [a, b] and $\mathbb{N} [\nu,\sigma]$ denotes the normal distribution with mean $\nu$ and variance $\sigma$. $N_f$ is grid frequency, which approximates the maximum possible frequency that can be resolved by the polynomials. For the flag, $\eta = 0$ means that the candidate stencil is smooth and should be selected for the final reconstruction, and vice versa. The final training database and the ANN network are open-source and can be downloaded from \url{https://github.com/Lilian6/NNENO}.
    \end{tablenotes}
\end{table}
\begin{itemize}
    \item No. 1: Linear functions, $k$ denotes the slope with a uniform distribution in the range of $[-5,5]$.
    
    \item No. 2: Smooth fluctuations with specific amplitudes and frequencies. $a_k$ denotes the amplitude, and $2\pi/k$ is the period. 
    
    \item No.~3: Exponential functions, the values of $a_1$ and $a_2$ are distributed uniformly in the range of $[-1,1]$.  
    
    \item No.~4: Triangular functions. $2P$ is the period, and $r/P$ denotes the amplitude. The values of $P$ are distributed uniformly in the range of $[2\Delta x, 8\Delta x]$.
    
    \item No.~5: Linear or constant profiles connected by kinks. 
    
    \item No.~6: Polynomials of different degrees connected by kinks. 
    
    \item No.~7: Elliptic functions. $\alpha$ and $a$ denote the height and the half width respectively. 
    
    \item No.~8: Step functions. Data are only collected at $x_{i+1/2} = x_0$, where the discontinuity is present.
    
    \item No.~9: Functions with high-frequency spectral property in analogy to Gibbs oscillations. 
   
\end{itemize}

\bibliographystyle{elsarticle-num-names}

\bibliography{explicit}

\end{document}